\documentclass{JHEP3}
\usepackage[english]{babel}
\usepackage{amssymb}
\usepackage{amsfonts}
\usepackage{amsmath}
\usepackage{graphicx}
\usepackage{epsfig}
\usepackage{cite}
\usepackage{psfrag}
\usepackage{subfigure}
\newcommand{\be}{\begin{equation}} \newcommand{\ee}{\end{equation}}
\newcommand{\bea}{\begin{eqnarray}} \newcommand{\eea}{\end{eqnarray}}
\newcommand{\beann}{\begin{eqnarray*}}  \newcommand{\eeann}{\end{eqnarray*}}
\newcommand{\bfig}{\begin{figure}} \newcommand{\efig}{\end{figure}}
\newcommand{\ba}{\begin{array}} \newcommand{\ea}{\end{array}}
\newcommand{\bcen}{\begin{center}} \newcommand{\ecen}{\end{center}}
\newcommand{\btab}{\begin{tabular}} \newcommand{\etab}{\end{tabular}}

     \def\sign{\operatorname{sign}}
\renewcommand{\Re}{\mathop{\rm Re}}   
\newtheorem{Proposition}{Proposition}[section]

\newtheorem{Theorem}{Theorem}[section]
\newtheorem{Lemma}{Lemma}[section]
\newtheorem{Corrolary}{Corrolary}[section]

\newcommand{\bp}{\begin{Proposition}}	\newcommand{\ep}{\end{Proposition}}
\newcommand{\bt}{\begin{Theorem}}	\newcommand{\et}{\end{Theorem}}
\newcommand{\bl}{\begin{Lemma}}		\newcommand{\el}{\end{Lemma}}
\newcommand{\bc}{\begin{Corrolary}}	\newcommand{\ec}{\end{Corrolary}}

\title{Evolution of Two-Point Functions from Holography}
\author{Jo\~ao Apar{\'\i}cio, Esperanza L\'opez\\
   Instituto de F\'{\i}sica Te\'orica CSIC/UAM\\
   Facultad de Ciencias, modulo C-8\\
   Universidad Aut\'onoma de Madrid\\
  28049 Madrid, Spain\\
  E-mail: \email{jpmn.aparicio@gmail.com, esperanza.lopez@uam.es}\\
}
\abstract{We consider a thermalization process in a 2-dimensional CFT that has a holographic description in terms of the gravitational collapse of a thin shell of null dust. This model represents a sudden perturbation of the CFT vacuum that communicates a uniform energy density to the system. We study the evolution of two-point functions at spacelike separated points $(t_1,l)$ and $(t_2,0)$, and reproduce the generic pattern first derived from the analysis of quantum quenches to critical systems. A crucial characteristic of these setups is that the excitations generated by the initial perturbation presents non-trivial quantum correlations.  As a consequence, for any $t_i\!<\!\infty$ equilibration is only effective on finite regions whose size grows as a lightfront. The behavior on larger regions is greatly determined by the initial state, which for the quenches we consider and the holographic model has relevant differences. However in both cases for late times the dependence on the scale $l$ of the two-point functions enters through the effective distance $l\!-\!t_1\!-\!t_2$. We interpret the onset of this behavior as an equilibration time for occupation numbers in these 2-dimensional models. 
}

\preprint{IFT-UAM/CSIC-11-59} 

\begin{document}

\section{\label{sec:intro}Introduction}

The AdS/CFT correspondence represents a conceptual breakthrough in the study of strongly coupled field theories. It has proven to be extremely valuable to obtain universal properties of field theories in the strongly coupled regime, although the concrete models accessible to construction are often not as realistic as desirable. The correspondence has been very successfully applied in recent years to the study of the transport properties of strongly coupled plasmas in the context of linear \cite{Son:2007vk} and non-linear \cite{Baier:2007ix,Bhattacharyya:2008jc} fluid dynamics. A natural continuation of this line of work, which is attracting a growing interest, is its application to the study of thermalization processes starting from a far from equilibrium state \cite{Chesler:2008hg,Beuf:2009cx,Chesler:2009cy,Bhattacharyya:2009uu,Das:2010yw,AbajoArrastia:2010yt,Albash:2010mv,Hashimoto:2010wv,Balasubramanian:2010ce,Erdmenger:2011jb,CaronHuot:2011dr,Balasubramanian:2011ur,Heller:2011ju,Garfinkle:2011hm}.  

The holographic dictionary relates the plasma phase of strongly coupled field theories at thermal equilibrium to black hole geometries \cite{Witten:1998zw}. Hence the holographic counterpart of a thermalization process must correspond to a process of gravitational collapse ending in the formation of a black hole \cite{Banks:1998dd,Danielsson:1999zt,Danielsson:1999fa,Giddings:2001ii}. The simplest observables which can be used as probes of the thermalization process are one-point functions, since they can be derived from an expansion of the dual background close to the boundary. In \cite{AbajoArrastia:2010yt,Albash:2010mv,Balasubramanian:2010ce,Balasubramanian:2011ur} it was initiated the study of the holographic evolution using observables which need information from the dual geometry far from the boundary. 

The dual geometry used in \cite{AbajoArrastia:2010yt,Albash:2010mv,Balasubramanian:2010ce,Balasubramanian:2011ur} was a Vaidya metric. These metrics describe the collapse of a shell of null dust to form a black hole, and are known analytically also for spacetimes with negative cosmological constant. This setup represents a CFT starting in its vacuum state and undergoing a translationally invariant perturbation, modeled by the collapsing shell, which brings it out of equilibrium. After the perturbation ceases, the system evolves according to the initial CFT hamiltonian. A similar dynamical setup is provided by a quantum quench, which denotes an action on a system in which a parameter of the hamiltonian is suddenly changed triggering the subsequent evolution. Holographic models for quantum quenches that affect some localized degrees of freedom have been proposed in \cite{Das:2010yw,Hashimoto:2010wv}.

Quantum quenches are unitary processes. As such, the excitations produced by the quench present non-trivial quantum correlations. This has very important consequences for the evolution towards equilibration of the system. In order that an observable involving an scale $l$ has reached its equilibrium value at a time $t$, it must happen that quantum entangled excitations have separated at least a distance $l$ at that time \cite{Calabrese:2005in}. Causality then implies that relaxation does not happen in the system globally, but it takes longer the bigger the region considered. Thermalization processes described holographically through gravitational collapse are also unitary \cite{AbajoArrastia:2010yt,Takayanagi:2010wp}. Using a 3-dimensional Vaidya metric, the dependence of the size of thermalized regions with time was reproduced in \cite{AbajoArrastia:2010yt} in perfect agreement with previous results \cite{Calabrese:2005in}. The observable used to follow the dynamical process in both works was the entanglement entropy.

There are some relevant differences between the Vaidya model of thermalization and quantum quenches. Quenches from a gapped to a critical system were analyzed at a general level in \cite{Calabrese:2005in,Calabrese:2006rx,Calabrese:2007rg}. This type of quenches differ from the holographic model in the entanglement pattern of the initial state that triggers the evolution. Namely in the former entanglement is initially localized on small scales, whereas in the Vaidya model there are long range correlations in the early stages of the evolution \cite{AbajoArrastia:2010yt}. A quantum quench between two different critical models was studied in \cite{cazalilla}, exhibiting power law correlations at late times after the quench. The holographic model has in common with it the existence of long range correlations in the initial state. However at late times it presents by construction a thermal behavior, in closer analogy with quenches from gapped models. 

The following remark is in order. Although in a unitary process information is not lost at the microscopical level, realistic perturbations in extended systems generically populate many energy levels and lead to relaxation towards an state that at the macroscopical level can be described as thermal \cite{reimann}. This is the case of the holographic models based on gravitational collapse. The final state towards which a system relaxes can however differ from thermal equilibrium if additional integrals of motion are present. This happens for example in the quench between two critical systems of \cite{cazalilla}, and in the quench from a massive to a massless free boson treated in \cite{Calabrese:2007rg}.

In \cite{Albash:2010mv} the holographic evolution of entanglement entropy was considered for boundary dimension $d\!=\!3$, with analogous results to \cite{AbajoArrastia:2010yt}. In addition to the entanglement entropy, equal-time two point functions and Wilson loops were studied in \cite{Balasubramanian:2010ce,Balasubramanian:2011ur} as probes of a thermalization process for $d=2,3,4$. The dynamical background they considered was the collapse of an infinitely thin shell of null dust, a limit of the Vaidya setup. For $d\!=\!2$, this simplified background allowed to obtain exact expressions instead of the numerical methods necessary in the smooth Vaidya case. 


The aim of this paper is to calculate general two-point functions in the thin shell geometry. We will compare our results with those of \cite{Calabrese:2006rx,Calabrese:2007rg}, where the evolution of one and two-point functions for quenches from gapped to gapless systems was analyzed, finding again perfect agreement. The general methods used in \cite{Calabrese:2005in,Calabrese:2006rx,Calabrese:2007rg} to study the dynamics of these quenches were based on the special properties of the conformal group in two-dimensions. They rely in performing the calculations on euclidean signature, and then prolonging to real time. The holographic methods allow for a direct calculation on lorentzian signature. Although we are using a very simple geometry to simulate a thermalization process, it is likely that our conclusions apply equally to more general models of thermalization based on gravitational collapse. 

Besides the influence of the initial pattern of quantum entanglement, a very important aspect of the evolution towards equilibration is the redistribution of energy among the excitations generated by the initial perturbation. It was suggested in \cite{AbajoArrastia:2010yt} the possibility of deriving from the holographic evolution of the entanglement entropy a time at which the redistribution of energy among modes according to thermal equilibrium should be nearly completed. An important result of the present work is to show that precisely the same threshold time is obtained from the analysis of general two-point functions.


The organization of the paper is as follows. Section 2 summarizes the results for the evolution of two-point functions after a quantum quench derived in \cite{Calabrese:2006rx,Calabrese:2007rg}. In Section 3 we recall the holographic evaluation of two-point functions for operators with high conformal dimension, that can be obtained in terms of the proper length of geodesics that anchor in the AdS boundary \cite{Banks:1998dd,Balasubramanian:1999zv}. We review the construction of geodesics in AdS$_3$ and in a BTZ black hole. The derivation of geodesics in the infalling shell geometry is addressed in Section 4. Sections 5 and 6 consider particular limits in which explicit expressions for the holographic two-point functions can be obtained: the case of large space separations for the two-point functions and that of late times after the perturbation respectively. Section 7 contains a discussion of our results. Several technical facts are collected in two appendices.

\section{Two-point functions after a quantum quench}

The two-point function of primary operators in the vacuum of an euclidean 2-dimensional CFT on the plane is completely fixed by symmetries. 
Since we will be interested in dynamical situations we need its prolongation to real time, with the result 
\be
\langle {\cal O}(t_1,l) {\cal O}(t_2,0)\rangle={1 \over (l^2-\Delta t^2)^\Delta} \, ,
\label{2pvac}
\ee
where $\Delta$ is the conformal dimension of the operator $\cal O$  and $\Delta t\!=\!t_1\!-\!t_2$. In the same way, conformal symmetry determines the two-point function at thermal equilibrium
\be
\langle {\cal O}(l,t_1) {\cal O}(0,t_2)\rangle=\left[ \left( {2 \pi \over \beta} \right)^2  \! {1 \over  2 \big( \! \cosh(2 \pi l/ \beta) - \cosh(2 \pi \Delta t / \beta) \big)} \right]^\Delta \, ,
\label{corr1}
\ee
where $\beta$ is the inverse of the temperature. Both in vacuum and at thermal equilibrium the two-point functions have singularities on the light-cone, meaning that in a 2-dimensional CFT all excitations propagate at the speed of light even in the presence of a thermal bath. In order to make precise the prolongation to real time of the euclidean propagator, a prescription has to be given that determines how to treat the lightcone singularities.
Depending on the prescription chosen, the Wightman function or the time-ordered two-point function can be obtained. The Wightman functions above require $\Delta t\!\rightarrow\! \Delta t-i 0^+$.

The evolution of one- and two-point functions after a quantum quench was analyzed in \cite{Calabrese:2006rx,Calabrese:2007rg}, following previous work on the entanglement entropy \cite{Calabrese:2005in}. As in that case, explicit results can be derived for 2-dimensional field theories when the system before the quench is in the ground state of a massive hamiltonian, while the dynamics after the quench is conformal. The initial ground state, which is not an eigenstate of the new hamiltonian, provides the starting point for the evolution. As customary, the quench is taken to happen at $t\!=\!0$.

In \cite{Calabrese:2005in,Calabrese:2006rx,Calabrese:2007rg} it was argued that the evolution after a quantum quench can be obtained from the continuation to real time of two-point functions in a strip geometry of broadness $2 \tau_0$, with $\tau_0$ a parameter of the order of the inverse mass gap in the initial state. The strip geometry can be mapped to the upper half-plane through the conformal transformation $\omega(z)\!=\!{2 \tau_0 \over \pi} \log z$, where $\omega$ and $z$ are the coordinates in the strip and  upper half-plane (UHP) respectively. Then 
\be
\langle {\cal O}(\tau_1,l) {\cal O}(\tau_2,0) \rangle= |\omega'(z_1)|^{-\Delta}  |\omega'(z_2)|^{-\Delta}\langle {\cal O}(z_1) {\cal O}(z_2) \rangle_{\rm UHP} \, ,
\label{2qq}
\ee
with $\omega(z_1)=l+i\tau_1$ and  $\omega(z_2)=i\tau_2$. The two-point function on UHP is well known \cite{Cardy:1984bb}
\be
\langle {\cal O}(z_1) {\cal O}(z_2) \rangle_{\rm UHP}  = \left( {z_{1 {\bar 2}} z_{2{\bar 1}} \over z_{12} z_{{\bar 1} {\bar 2}} z_{1{\bar 1}} z_{2 {\bar 2}} }  \right)^\Delta
F\left( {z_{1{\bar 1}} z_{2 {\bar 2}} \over z_{1 {\bar 2}} z_{2{\bar 1}}} \right) \, ,
\ee
where $z_{ij}\!=\!|z_i\!-\!z_j|$ and the function $F$ depends on the particular CFT and boundary conditions under consideration. Rotating to real time, $\tau_i\!=\! \tau_0\!+\!it_i$, the model independent piece of the two-point function \eqref{2qq} is
 \be
 \left[ \left({\pi \over 2 \tau_0} \right)^2 { \cosh(\pi l/2 \tau_0)+\cosh(\pi(t_1+t_2)/2\tau_0) \over 4 \cosh(\pi t_1/2 \tau_0) \cosh(\pi t_2/2 \tau_0) \big(\! \cosh(\pi l /2\tau_0)-\cosh(\pi(t_1-t_2)/2 \tau_0)\big)} \right]^{\Delta} \, .
 \label{cp}
 \ee
Taking moreover the limit $t_i$, $l\!\gg\!\tau_0$, the argument of the function $F$ gives 
\be
 {z_{1{\bar 1}} z_{2 {\bar 2}} \over z_{1 {\bar 2}} z_{2{\bar 1}}} \simeq {e^{\pi(t_1+t_2)/2\tau_0} \over e^{\pi l/ 2 \tau_0}+e^{\pi(t_1+t_2)/2\tau_0} } \, .
 \label{arg}
 \ee

When $l\!<\!t_1\!+\!t_2$ the argument of $F$ is approximately one, and for any CFT it is verified $F(1)\!=\!1$ \cite{Calabrese:2006rx,Calabrese:2007rg}. In this case, the contribution \eqref{cp}  reproduces the result for the two-point function at thermal equilibrium \eqref{corr1} with inverse temperature $\beta = 4\tau_0$. This leads to interpret $1/4 \tau_0$ as an effective temperature for the long wavelength modes. This same conclusion was obtained before from the evolution of the entanglement entropy \cite{Calabrese:2005in}.

When instead $l\!>\!t_1\!+\!t_2$, the argument of $F$ is very small, of order $e^{-\pi(l-t_1-t_2)/ 2 \tau_0}$. Remarkably, also in this limit $F$ has a generic form  \cite{Cardy:1984bb,Calabrese:2006rx,Calabrese:2007rg}
\be
F(z) \propto z^x \, .
\ee
The exponent $x$ depends both on the operator ${\cal O}$ and the initial state for the evolution, and it is only different from zero if ${\cal O}$ has a vanishing one-point function. In that case, and assuming again large times and interval we have
\be  
\langle {\cal O}(l,t_1) {\cal O}(0,t_2)\rangle \propto e^{-\Delta \pi (t_1+t_2)/2 \tau_0} e^{-x \pi  (l-t_1-t_2)/2 \tau_0} \, .
\label{noneq2p}
\ee
The transition region between both regimes happens at $l\!\sim\!t_1\!+\!t_2$ over a region of order $\tau_0$. It is governed by the function $F$ and thus model dependent.

The physical picture that explains the previous evolution pattern is as follows. The non-zero energy density created at t = 0 by the quench translates into the production of a translationally invariant sea of excitations. As the quench is a unitary process, there will be quantum correlation between the resulting excitations.
Entangled excitations will tend to spread over ever larger regions. In order that observables inside a region can behave as thermalized, the distance between left and right moving entangled excitations must be bigger than the size of the region. In this way each component of an entangled pair can act as part of a thermal bath for the other. A simple picture where entangled excitations are assumed to move classically and without scattering, was further proposed in \cite{Calabrese:2005in}. For the critical quench described above, all excitations are taken to propagate at the speed of light. Since the ground state before the quench has a mass gap and hence a finite correlation length, only excitations produced approximately at the same point will be entangled. Both facts give rise to a sharp horizon effect. In order that a region of size $l$ behaves as thermalized, a time $l/2$ after the quench is required. Moreover a variation in the size of a region of the order of the initial correlation length around $l\!=\!2t$ will bring from regions that behave as thermalized to those that contain the main information about the entanglement pattern of the initial state. In this second case, the expectation value of  observables should be reminiscent of that in the initial state. This is indeed what happens for the entanglement entropy, which is defined as the von Neumann entropy of the reduced density matrix associated to a subregion of the system. Its evolution for the case of a single interval was studied in \cite{Calabrese:2005in}.

The same picture explains the behavior of non-equal time two-point functions \cite{Calabrese:2006rx,Calabrese:2007rg}. The distance that separates a pair of left and right moving entangled excitations produced at $t\!=\!0$ and propagating up to generic different times is $t_1\!+\!t_2$. When $l\!<\!t_1\!+\!t_2$ there are no entanglement excitations over the distance involved by the two-point function and a thermal result is reproduced. On the contrary, when $l\!>\!t_1\!+\!t_2$ there are entangled excitations on distances of order $l$.  
Their pattern of entanglement is determined by the initial state, while the effective distance their quantum correlations feel is $l\!-\!t_1\!-\!t_2$. 
The fact that their contribution to the two-point function \eqref{noneq2p} is exponentially suppressed can be traced back to the fact that the initial state is massive. Notice that $x/\tau_0\!\sim \!x m$, where $m$ is the mass gap in the initial state and $x$ is a parameter containing information about the initial state. The two-point function \eqref{noneq2p} has also an exponential suppression with $t_1\!+\!t_2$. This combination gives a measure of the decoherence due to the growing separation of entangled excitations. This fact is independent of the initial state. Hence the only information about the initial state involved in it is $\tau_0$, which sets the energy density generated by the quench.

\section{\label{sec:geostatic}Holographic two-point functions}

According to the AdS/CFT dictionary, two-point functions of gauge invariant operators are derived from the on-shell supergravity action for the bulk fields that couple to the chosen operators \cite{Gubser:1998bc,Witten:1998qj}. This requires to know the bulk to boundary propagator associated to that field, which can only be obtained analytically in very simple examples. In particular, for a time-dependent background the derivation of the bulk to boundary propagator will involve solving numerically a partial differential equation. 

An alternative way to derive boundary Green's functions was proposed in \cite{Banks:1998dd}. They should be given in terms of bulk correlators for the associated fields to which the operators couple. Bulk propagators can be represented as an integral over paths that join the insertion points $(t_1,l;r_b)$ and $(t_2,0;r_b)$, having then
\be
\langle {\cal O}(t_1,l) {\cal O}(t_2,0)\rangle =\lim_{r_b \rightarrow \infty} r_b^{2\Delta} \int  {\cal D}{\cal P} e^{- \Delta L({\cal P})} \, ,
\label{path2p}
\ee
where $L({\cal P})$ is the proper length of the path with the convention to be real for space-like trajectories. For operators with very high conformal dimension $\Delta$ it is justified to perform a saddle point approximation, which reduces the path integral to a sum over geodesics \cite{Balasubramanian:1999zv}. It should be stressed that this approximation requires special care in lorentzian spaces because of subtleties associated with the choice of steepest descendent contour \cite{Louko:2000tp,Fidkowski:2003nf,Festuccia:2005pi}\footnote{It was shown in \cite{Fidkowski:2003nf}  that for Schwarzchild-AdS$_d$, with $d\!>\!3$, there exist 2-point functions receiving contributions from several geodesics. In some cases the largest contribution comes not from a real but from a complexified geodesic. 
This analysis requires the geometry of the space under consideration to admit an analytic continuation, which is not the case of the infinitely thin shell background.}.
We will assume that these complications do not arise in the 3-dimensional collapsing shell background we will be interested in, as it was also done in \cite{Balasubramanian:2010ce,Balasubramanian:2011ur}. The consistency of the results we will serve as a check of this assumption.

The geodesic length has a divergent logarithmic contribution from the region $r_b \!\rightarrow\!\infty$. A regularized geodesic length can be defined by subtracting this piece
\be
L=L_{reg}+2 \log r_b \, .
\label{lreg}
\ee
The logarithmic contribution to $L$ precisely cancels the power of $r_b$ in \eqref{path2p} and the limit in this expression can be easily taken. Relation \eqref{path2p} reduces then to evaluating the regularized length of geodesics with endpoints in the boundary. In the 3-dimensional case we are considering, there is a unique geodesic that joins two given boundary points. Hence the two-point function of operators with high conformal dimension will be simply given by
\be
\langle {\cal O}(t_1,l) {\cal O}(t_2,0)\rangle = e^{-\Delta L_{reg}(t_1,l;t_2,0)} \, .
\label{hol2p}
\ee

Two-point functions are not the only observable whose holographic determination involves evaluating the length of bulk geodesics. This is also the case of the entanglement entropy \cite{Ryu:2006bv,Ryu:2006ef}. The entanglement entropy of a single interval in a 2-dimensional CFT is given holographically by the length of the equal-time geodesic that anchors on the boundary at the interval endpoints. The study of equal-time geodesics in a Vaidya geometry representing the collapse of a null dust shell with finite thickness was carried out numerically in \cite{AbajoArrastia:2010yt}. The limit of an infinitely thin shell in AdS$_3$ allows to use analytic instead of numerical methods. Equal-time geodesics were reconsidered in \cite{Balasubramanian:2010ce,Balasubramanian:2011ur} in this background. We want to extend this study to general geodesics with endpoints in the boundary. Using the approximation \eqref{hol2p}, we will reproduce the pattern for the evolution of two-point functions presented in the previous section. 

The geometry describing the infinitely thin shell is that of a BTZ black hole outside the shell and AdS$_3$ inside. Both spaces join along a radial null surface that we take to start at the boundary at $t\!=\!0$ in analogy with the convention for the quantum quench. This geometry represents an instantaneous action on the dual CFT vacuum that brings the system out of equilibrium by communicating an homogeneous energy density. The system will evolve subsequently towards equilibration at a temperature set by that of the dual black hole formed in the gravitational collapse process. In this set up, the only field theory observable with a non-zero one-point function is the stress tensor. The operators ${\cal O}$ for which we will compute the two-point functions using the geodesic approximation have thus vanishing one-point functions, as it was assumed in the quantum quench result \eqref{noneq2p}.

We will review now the description of geodesics in the AdS$_3$ and BTZ geometries, since we will need them to evaluate two-point functions in the thin shell background.
The metric of the non-compact BTZ black hole is given by
\be
ds^2=-\big( r^2-m \big) dt^2+ {dr^2 \over  r^2-m}+r^2 dx^2 \, ,
\label{BTZv}
\ee
where the parameter $m$ represents the black hole mass. The associated Hawking temperature is $T\!=\!\sqrt{m}/2\pi$ and the radius of the event horizon $r_h\!=\!\sqrt{m}$.
The case $m\!=\!0$  of the previous metric describes the AdS$_3$ solution in the Poincar\'e patch.
In the affine parameterization, the geodesic equations are
\begin{eqnarray}
\dot t &=& \frac{A}{r^2-m} \label{geo1}\\
\dot r &=& \pm \frac{1}{r}\sqrt{A^2r^2+(r^2-m)\left(Br^2-r_\star^2\right)} \label{B}\\
\dot x &=& \frac{r_\star}{r^2} \label{geo3}
\end{eqnarray}
where $r_\ast$, $A$ are integrals of motion associated with the Killing vectors $\partial_x$ and $\partial_t$, and therefore having an interpretation of conserved momentum and energy. The previous equations imply ${\dot x}^\mu {\dot x}^\nu\!g_{\mu \nu}\!=\!B$. If we take $\lambda$ to represent the geodesic proper length we have $B\!=\!1,0,-1$ for spacelike, lightlike and timelike geodesics respectively. Equation \eqref{B} implies that only when $B\!=\!1$ the associated geodesics can reach the boundary. Since this is what we need for the holographic calculation of the CFT two-point functions, from now on we will focus on spacelike geodesics in the bulk.

It will be convenient to describe the geodesics as functions $t(r),x(r)$ of the radial coordinate, instead of using the affine parameterization. As it is clear from \eqref{B} each function will have two branches, which we will denote as $\pm$. We choose them to be defined by the conditions
\be
t_+(\infty)\geq t_-(\infty) \, , \hspace{1cm} x_+(\infty) \geq x_-(\infty) \, .
\label{branchh}
\ee
The integration constant $r_\ast$ and $A$ can have either sign. The reason for the criterium \eqref{branchh} is that the resulting functions $t_\pm(r)$ and $x_\pm(r)$ depend only on the modulus of the integration constants. Hence without loss of generality we can consider $r_\ast$ and $A$ to be always positive, and shift the information about their sign to the choice of branch. Since the signs of both constants are independent, so will it be the choice of branch for the functions $t$ and $x$.

\subsection{\label{sec:AdS} AdS$_3$}

The geodesic equations \eqref{geo1}-\eqref{geo3} are straightforward to integrate for AdS$_3$. The result for spacelike geodesics is
\be
t_\pm(r)=t_0\pm t_{AdS}(r) \, , \hspace{1cm} x_\pm(r)=x_0\pm x_{AdS}(r) \, ,
\label{functionsAdS2}
\ee
where
\be
{t_{AdS} \over A}={x_{AdS} \over r_\ast}={\sqrt{A^2+r^2-r_\ast^2} \over r |r_\ast^2-A^2|}\, .
\label{functionsAdS}
\ee
Given the translational invariance in $t$ and $x$ of the geometry, the integration constants $t_0$ and $x_0$ do not have any influence on the geodesic length.
The separation between the geodesic endpoints $(t_1,x_1)$ and $(t_2,x_2)$, is determined by the constants of motion $r_\ast$ and $A$ as  
\be
|\Delta t|={2 A \over |r^2_\ast-A^2|} \; , \;\;\;\;\; l={2 r_\ast \over | r^2_\ast-A^2|} \, ,
\label{tlAdS}
\ee
where $\Delta t\!=\!t_1\!-\!t_2$ and $l\!=\!|x_1\!-\!x_2|$.

When $r_\ast\!>\!A$, the geodesics extend from the boundary to the value of the radial coordinate
\be
r_m=\sqrt{r^2_\ast-A^2} \, ,
\label{rmsp}
\ee
where both branches join. As can be seen from \eqref{tlAdS}, the geodesic endpoints are spacelike separated in this case. The lightlike case can be reached from the spacelike one in the limit $r_\ast\!\rightarrow\!A$. Keeping the separation between the geodesic endpoints \eqref{tlAdS} finite in this limit requires
\be
r_\ast=\rho \, , \hspace{1cm}  A=\rho-1/l \, ,
\ee
with $\rho\!\rightarrow\!\infty$. The minimal value of the radial coordinate $r_m$ tends then to infinity. This shows that geodesics joining two lightlike separated points in the case of a non-compact boundary, are lightlike geodesics that do not leave the boundary. 

When instead $r_\ast\!<\!A$  the branches $\pm$ describe disconnected trajectories that reach down to $r\!=\!0$, where $t,x\!=\!\pm\infty$. In this case there is no reason to take the integration constants $t_0$ and $x_0$ in \eqref{functionsAdS2} equal for both branches. Let us however do it, considering both trajectories as part of a single geodesic. The separation between its endpoints is again given by \eqref{tlAdS}, with the result that they are timelike separated. 

The proper length of the geodesic is
\be
L=2 \int^{r_b}_{r_0} {dr \over {\dot r}}=2 \int^{r_b}_{r_0} {d r \over \sqrt{A^2+r^2-r_\ast^2} }\, ,
\label{properAdS}
\ee
with the lower integration limit equal to $r_m$ in \eqref{rmsp} for $r_\ast\!>\!A$ and zero otherwise. The upper integration limit $r_b$ has been introduced in order to avoid the logarithmic divergence from the integration region close to the boundary, as explained above. The result for the regularized geodesic length \eqref{lreg}, both in the spacelike and timelike cases, is
\be
L_{\text{reg}}=- \log {|r^2_\ast-A^2| \over 4}=\log|l^2-\Delta t^2| \, .
\label{lads}
\ee
Using the geodesic approximation \eqref{hol2p}, the vacuum two-point function of primary operators with spacelike separation is reproduced \cite{Balasubramanian:1999zv}. 

It is tempting to apply the geodesic approximation to two-point functions with timelike separations. The length of the disconnected geodesics proposed above gives almost the right result. Substituting in \eqref{hol2p}, it reproduces the correct value for the two-point function up to a constant phase. This mismatch comes from the modulus in the argument of the logarithm, which for timelike separations is relevant. The presence of the modulus is actually unavoidable because with the convention we have chosen the proper length of  spacelike geodesics is always a real number. It would be very interesting to understand whether the geodesic approximation to the holographic derivation of two-point functions can exactly reproduce the timelike case, explaining the mismatch in a phase that we have found. It would be also important to determine whether the geodesic approximation can represent the different prescriptions for treating the lightcone singularities, leading to the Wightman or time-ordered two-point functions.

\subsection{\label{sec:BTZ}BTZ black hole}

The geodesic equations \eqref{geo1}-\eqref{geo3} in a BTZ background can be also explicitly integrated, obtaining
\be
\begin{array}{lll}
t_\pm (r) & =& \displaystyle t_0-{1 \over 2 r_h} \log \left[ m^{-1}  \Big(\epsilon X_t(r) \mp \sign(r^2\!-\!m) \sqrt{X_t^2(r)-C} \Big)\right]\,  ,  \\[4mm]  
x_\pm (r)  & =& \displaystyle x_0-{1 \over 2 r_h} \log \left[ m^{-1} \Big({\tilde \epsilon} X_x(r) \mp  \sqrt{X_x^2(r)-C}\Big) \right] \, ,
\label{functions}
\end{array}
\ee
where we have defined
\be
\begin{array}{c}
\displaystyle X_t(r)=r_\ast^2-A^2-m -{2 A^2 m\over r^2-m}\,\, , \hspace{.5cm} X_x(r)=r_\ast^2-A^2+m -{2 r_\ast^2 m \over r^2}  \, , \\[5mm]
C=  (r_\ast^2-A^2-m)^2-4 A^2 m= (r_\ast^2-A^2+m)^2-4 r_\ast^2 m\, ,
\end{array}
\label{CC}
\ee
together with 
\be
\epsilon = \sign (r_\ast^2-A^2-m) \, , \hspace{.8cm} {\tilde \epsilon}=\sign (r_\ast^2-A^2+m) \, .
\label{ee}
\ee

The value of the radial coordinate ranges from infinity to the zero of the square root in the functions \eqref{functions} with larger $r$ \footnote{As it should be, the same value is derived from $t(r)$ and $x(r)$.}
\be
r_m^2={1 \over 2} \big(r_\ast^2-A^2+m +\sqrt{C}\, \big) \, ,
\label{rmin}
\ee
whenever this quantity is real and positive. Otherwise we have two separated trajectories, both reaching the black hole singularity $r\!=\!0$. A necessary condition for this not to happen is that $C$ is positive, which we will assume in this subsection. This is the only case relevant for the geodesic approximation to the holographic derivation of two-point functions at thermal equilibrium. 

\begin{figure}[thbp]
\centering
\begin{tabular}{c}
\includegraphics[width=14cm]{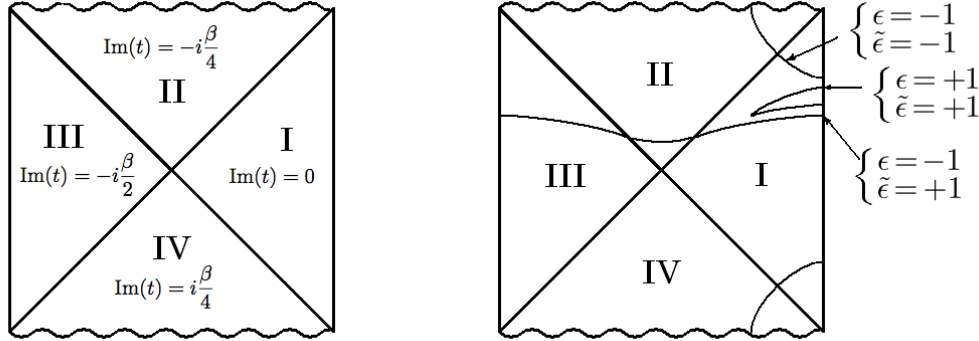}
\end{tabular}
\caption{\label{fig:btz} On the left we show the imaginary part of $t(r)$ in the different wedges of the Penrose diagram of the BTZ black hole. On the right we draw qualitatively different types of BTZ geodesics according to our classification.}
\end{figure}

In order to describe the geodesics we will need to consider the extended Penrose diagram of the BTZ black hole, which has a second asymptotic region (wedge III) and a past singularity (wedge IV). The time $t$ acquires a non-vanishing imaginary part in three of the four wedges that compose the extended diagram, as shown in Fig.\ref{fig:btz} \cite{Maldacena:2001kr,Kraus:2002iv}. 

The different geodesic types can be classified by the signs $\epsilon$ and ${\tilde \epsilon}$ \eqref{ee}, see also \cite{Balasubramanian:2011ur}. When $\epsilon\!=\!{\tilde \epsilon}\!=\!1$ the minimal value of the radial coordinate satisfies $r_m\!>\!\ r_h$, implying that these geodesics stay always outside the event horizon and are contained in wedge I or III. When $\epsilon\!=\!-1$ and ${\tilde \epsilon}\!=\!1$ the associated geodesics reach behind the event horizon but they do not fall into the singularity, since for them $0\!<\!r_m\!<\!\ r_h$. Close to the horizon the two branches of the geodesic time coordinate behave as 
\be
t_\pm(r) \simeq  \mp{1\over 2 r_h}\log(r\!-\!r_h) \, .
\ee
If we choose the branch $+$ to start at the boundary of wedge I, the branch $-$ must start at the boundary of wedge III in order that they can join at $r_m$ in wedge II \footnote{As already noticed, the integration constants do not need to coincide for both branches. In this case we should choose $t_{0-}\!=\!t_{0+}\!-\!i\beta/2$, with $t_{0+}$ real.}. Hence these geodesics connect the two separated AdS boundary regions of the eternal black hole (see Fig.\ref{fig:btz}). Finally when $\epsilon\!=\!{\tilde \epsilon}\!=\!-1$, we have two disjoint trajectories each of them reaching the singularity. As we did for AdS, let us however consider them formally as part of a single geodesic characterized by a common value of the integration constants $t_0$ and $x_0$. If branch $+$ starts on the boundary of wedge I and falls to the future singularity, then branch $-$ represents a trajectory that exits the past singularity on wedge IV and returns to wedge I.
   
The conserved quantities $A$ and $r_\ast$ determine the separation between the endpoints for the three types of geodesics described above
\be
\cosh {2 \pi \Delta t \over \beta}= {| r_\ast^2-A^2-m| \over \sqrt{C}} \, ,  \hspace{.7cm}   \cosh{2 \pi l \over \beta}={| r_\ast^2-A^2+m | \over \sqrt{C}} \, ,
\label{tl}
\ee
where for geodesics ending on opposite boundaries we have defined $\Delta t\!=\!\Re(t_1\!-\!t_2)$.
When the endpoints lie on the same boundary, their separation is spacelike (timelike) when $\epsilon\!=\!{\tilde \epsilon}\!=\!1(-1)$. For geodesics connecting both AdS boundaries, there is no restriction on the values of $\Delta t$ and $l$.
 
The proper length of the geodesic in the BTZ background is
\be
L=2 \int^{r_\infty}_{r_m} {dr \over {\dot r}}=\int^{r^2_b}_{r^2_m} {d r^2 \over \sqrt{A^2r^2+(r^2-m)(r^2-r_\ast^2)} }\, .
\label{proper}
\ee
The integration can be explicitly performed, with the result for the regularized length
\be
L_{\text{reg}}=- \log {\sqrt{C} \over 4}\, .
\label{lprop}
\ee 
For geodesics with $\epsilon\!=\!{\tilde \epsilon}\!=\!-1$, $r^2_m$ \eqref{rmin} is negative. However the integral in the second equality of \eqref{proper} can be generalized to this case by analytically prolonging its integrand to negative values of the radial coordinate square $r^2\!\geq \! r_m^2$. Remarkably, this leads to the same result \eqref{lprop} for the geodesic length.

Using \eqref{tl} to express $C$ in terms of $l$ and $\Delta t$, and substituting in \eqref{hol2p} we obtain 
\be
\langle {\cal O}(l,t_1) {\cal O}(0,t_2)\rangle=\left[ \left( {2 \pi \over \beta} \right)^2  \! {1 \over  2 \,| \! \cosh(2 \pi l/ \beta) - \epsilon {\tilde \epsilon}  \cosh(2 \pi \Delta t / \beta) |} \right]^\Delta \, .
\label{corr}
\ee
When $\epsilon\!=\! {\tilde \epsilon} \!=\!1$, \eqref{corr} reproduces the two-point function at thermal equilibrium for spacelike separations in a 2-dimensional CFT \cite{Louko:2000tp}. As it was the case in AdS$_3$, $\epsilon\!=\! {\tilde \epsilon} \!=\!-1$ gives the right answer for two-point functions with a timelike separation up to a phase.

It has been proposed that geodesics connecting the two asymptotic AdS boundaries of the eternal black hole are associated with field theory correlators in the Schwinger-Keldysh, or real-time formalism \cite{Maldacena:2001kr,Herzog:2002pc}. In this formalism the tensor product of two copies of the original field theory is considered. The following pure but entangled state is associated with the system at thermal equilibrium
\be
| \Psi\rangle={1 \over Z^{1/2}} \sum_i e^{-{1 \over 2} \beta E_i} |E_i\rangle_1 \otimes |E_i \rangle_2 \, ,
\label{skst}
\ee
with $|E_i\rangle$ energy eigenstates of the theory. Ordinary thermal correlators are obtained when all operators are inserted on the same copy of the system. In the state \eqref{skst}, correlators with operators acting on different copies of the system can be related with ordinary ones by
\be
\langle \Psi| {\cal O}_1(l,t_1) {\cal O}_2(0,t_2)|\Psi\rangle \rangle =\langle \Psi|{\cal O}_1(l,t_1) {\cal O}_1(0,-t_2-i\beta/2)|\Psi \rangle \, .
\ee
The holographic 2-point function \eqref{corr} agrees with this relation. Using the bulk to boundary propagator, two-point function between operators inserted on both copies have
been obtained in the context of the AdS/CFT correspondence in \cite{Maldacena:2001kr,Kraus:2002iv}.

\section{\label{sec:geodyn}Geodesics in the infalling shell geometry}

Our aim is to construct general geodesics with endpoints on the boundary in the dynamical geometry describing the collapse of an infinitely thin shell of null dust. In order to describe both the metric and the geodesics it is convenient to substitute the time coordinate $t$, which is not constant across the shell, by an infalling radial null coordinate $v$ which it is. Since we are taking the shell to start its collapse at $t\!=\!0$ from the boundary, its trajectory will be given by $v\!=\!0$. Hence the associated metric is
\be
ds^2=-\big( r^2-m(v) \big) dv^2+2 dr dv+r^2 dx^2 \, ,
\label{Vaidya}
\ee
with $m(v)\!=\!0$ for $v\!<\!0$, describing empty AdS$_3$, and $m(v)\!=\!m$ for $v\!>\!0$, describing a BTZ black hole. The more general Vaidya metrics correspond to consider an arbitrary mass function $m(v)$. The relation between $v$ and $t$ in the AdS and BTZ backgrounds is
\begin{eqnarray}
v & = & t -1/r \, , \hspace{35mm}   {\rm AdS} \, , \label{vads} \\
v & = & t+{1 \over 2 r_h} \log {|r-r_h| \over r+r_h} \, , \hspace{15mm}  {\rm BTZ} \, . \label{vbtz}
\end{eqnarray}

We will assume that the geodesics endpoints $(t_1,x_1)$ and $(t_2,x_2)$ satisfy 
\be
t_1>t_2 \, , \hspace{1cm} x_1-x_2=l>0 \, .
\label{bbcc}
\ee 
This choice does not represent a loss of generality since parity, which is a symmetry of our set up, changes  \eqref{bbcc} to $x_1\!<\!x_2$.
From the criterium \eqref{branchh}, the previous choice implies
\be
v_+(\infty)=t_1\; , \, v_-(\infty)=t_2 \; , \hspace{1cm}  x_+(\infty)=x_1\; , \, x_-(\infty)=x_2 \, .
\label{bbcc1}
\ee 

There are two types of geodesics to consider. 
We have those that stay entirely outside the shell and only feel the BTZ part of the geometry. These geodesics reproduce the result for two-point functions at thermal equilibrium. 
Then there are geodesics that intersect the trajectory of the shell. Actually no matter how large $t_1$ and $t_2$ are taken after the perturbation in the case with a non-compact boundary, there will always be an $l$ such that the associated geodesic reaches the shell \cite{AbajoArrastia:2010yt}. Indeed the larger $l$ is, the deeper the associated geodesics will get in the black hole geometry and in particular the closer to the horizon $r_h$. The second term in the rhs of \eqref{vbtz} makes then a dominant contribution that brings the null coordinate $v$ towards small values, untill $v\!=\!0$ is attained. Geodesics with larger values of $l$ will intersect the trajectory of the shell and lead to a non-equilibrium result for the two-point functions.

\subsection{Thermal geodesics}

The condition for geodesics to lie in the BTZ part of the geometry was studied in \cite{AbajoArrastia:2010yt} for the equal time case $t_1\!=\!t_2\!=\!t$. The following very simple relation turns out to hold for equal time BTZ geodesics that anchor on the same AdS boundary
\be
l=2(t-v_\ast) \, ,
\label{BTZgeoet}
\ee
where $v_\ast$ denotes the minimal value of the coordinate $v$ attained by the geodesic. Since the shell follows the trajectory $v\!=\!0$, geodesics with $l\!<\!2t$ will lead to a thermal result for the associated observables: entanglement entropy and equal-time two-point functions. This reproduces the field theory analysis of quantum quenches reviewed in Section 2. Namely, in order that a region of size $l$ can appear as thermalized, a time at least equal to $l/2$ has to be waited \cite{Calabrese:2005in}. 

\begin{figure}[thbp]
\centering
\begin{tabular}{c}
\includegraphics[width=7.2cm,height=4cm]{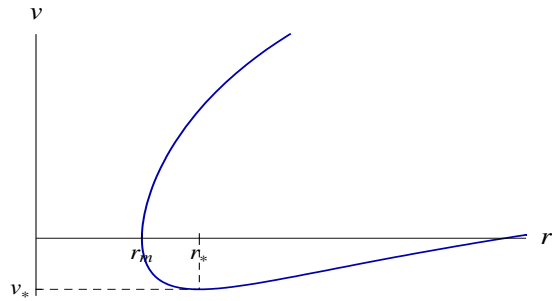}
\end{tabular}
\caption{\label{fig:exampleBTZ} Example of a non equal-time BTZ geodesic completely contained in wedge I, {\it i.e.} with $\epsilon\!=\!{\tilde \epsilon}\!=\!1$ and $C$ \eqref{CC} positive. AdS$_3$ geodesics with $r_\ast\!>\!A$ verify the same pattern.}
\end{figure}

As we also saw in Section 2, the threshold for thermal behavior of two-point functions involving different times after a quench was $l\!=\!t_1\!+\!t_2$ with $t_i\!>\!0$. Thus at threshold the separation between the insertion points of the two-point function is spacelike. In order to reproduce this bound holographically it is then enough to focus on geodesics with spacelike separated endpoints, {\it i.e.} those contained in wedge I. Equal-time geodesics are characterized by having a vanishing energy parameter $A$, such that  the functions $v_\pm(r)$ for them coincide. The minimal value of the radial coordinate that they can attain coincides with $r_\ast$ and by symmetry at this point $v_\ast$ is  reached. For geodesics with endpoints at different times the two branches $v_\pm(r)$ are different, joining smoothly at the minimal value of the radial coordinate $r_m$  \eqref{rmin}, which now differs from $r_\ast$ (see Fig \ref{fig:exampleBTZ}).
Given the criterium \eqref{branchh} chosen to distinguish between branches, it is clear that the minimum value of the null coordinate must occur in the branch $-$. Thus we should search for a zero of
\be
{d v_- \over d r} = {1 \over r^2-m} \left( 1-{A r \over \sqrt{A^2 r^2 +(r^2-m)(r^2-r_\ast^2)}} \right) \, ,
\label{dvdr}
\ee
where according to our conventions $A$ is always positive. This expression has a zero at infinity, corresponding to a maximum. It has an additional zero at $r_\ast$, necessarily then a minimum, implying $v_\ast\!=\!v_-(r_\ast)$. This discussion applies as well to non equal-time geodesics in AdS$_3$ with $r_\ast\!>\!A$.

Substituting in \eqref{vbtz} and \eqref{functions} leads to
\be
v_\ast=t_0-{1 \over 2 r_h} \log\Big[m^{-1}  \big( (r_\ast+r_h)^2-A^2 \big) \Big] \, ,
\ee
The value of the constant $t_0$ is set by the boundary conditions \eqref{bbcc1}
\be
t_0={t_1+t_2 \over 2} + {1 \over 4 r_h} \log \big( C m^{-2} \big) \, , 
\ee
and we then have
\be
v_\ast={t_1+t_2 \over 2}+{1 \over 4 r_h} \log { (r_\ast-r_h)^2-A^2 \over  (r_\ast+r_h)^2-A^2} \, .
\ee
Inverting the relation between $l$ and the integrals of motion $A$ and $r_\ast$ \eqref{tl}, we observe that the second term on the rhs equals $-l/2$.
Hence we finally obtain
\be
l=t_1+t_2-2 v_\ast \, ,
\label{geogen}
\ee
which generalizes \eqref{BTZgeoet} to geodesics with arbitrary spacelike separated endpoints. For an infalling shell with trajectory $v\!=\!0$, representing a sudden perturbation of the dual CFT at $t\!=\!0$, we reproduce the same threshold for thermal behavior as that found in the study of quantum quenches \cite{Calabrese:2006rx,Calabrese:2007rg}.

\subsection{Non-equilibrium geodesics}

We turn now to the construction of geodesics that intersect the infalling shell. The matching conditions for geodesics across the shell have been derived in \cite{Balasubramanian:2011ur}. Let us briefly reobtain them here for completeness. Using the affine parameterization, the geodesic equations are
\begin{eqnarray}
\dot x &=&  {r_\star \over r^2}  \, , \label{xdot} \\
\ddot v &=&  -r \dot v^2 + {r_\star^2 \over r^3} \, , \label{vdot} \\
\ddot r & = &  2 r {\dot r} {\dot v} +(r^2-m \Theta(v)) {\ddot v} -{1 \over 2} m \delta(v) {\dot v}^2 \, , \label{rdot}
\end{eqnarray}  
where $\Theta(v)$ is the step function. In the AdS and BTZ backgrounds the geodesic equations contained two integration constants, $r_\ast$ and $A$, associated to invariance under $x$ and $t$ translations. The collapse geometry only preserves invariance under translations in $x$, hence above appears only the associated conserved quantity $r_\ast$. The breakdown of time translations implies that the energy parameters in the AdS and BTZ segments of geodesics that intersect the shell will differ.

Let us call $\lambda_c$ the value of the affine parameter at an intersection point between the geodesic and the infalling shell. We will denote with a subindex "out" the segment that lives outside the shell, perceiving a BTZ geometry, and use a subindex "in" for the segment inside the shell, feeling AdS$_3$. Continuity requires that 
\be
x_\text{in}(\lambda_c)=x_\text{out}(\lambda_c) \, , \;\;\;\;\; v_\text{in}(\lambda_c)=v_\text{out}(\lambda_c) \, , \;\;\;\;\; r_\text{in}(\lambda_c)=r_\text{out}(\lambda_c) \, .
\label{confun}
\ee 
Then \eqref{xdot} immediately implies 
\be
\dot x_\text{in}(\lambda_c) = \dot x_\text{out}(\lambda_c)  \, .
\label{con1}
\ee

The behavior of $\dot v$ and $\dot r$ across the shell can be obtained by integrating the rhs of the geodesic equations between $\lambda_+\! > \!\lambda_c\!>\!\lambda_-$. 
Unless the integrand contains at least delta function singularity, letting $\lambda_\pm\!\rightarrow\!\lambda_c$ will lead to a vanishing result which would imply the continuity of the associated first derivative. Notice that a delta function can not appear in the first derivative of a continuous function. Applying this reasoning to \eqref{vdot} we obtain
\be
\dot v_\text{in}(\lambda_c) = \dot v_\text{out}(\lambda_c) \, .
\label{con2}
\ee
However the first derivative of the radial coordinate acquires a discontinuity due to the last term on \eqref{rdot}
\be
\dot r_\text{out}(\lambda_c) - \dot r_\text{in}(\lambda_c) = -\frac{1}{2} m \dot v(\lambda_c) \, .
\label{con3}
\ee
 
\begin{figure}[thbp]
\centering
\begin{tabular}{cc}
\subfigure[]{\includegraphics[width=6.8cm,height=3.3cm]{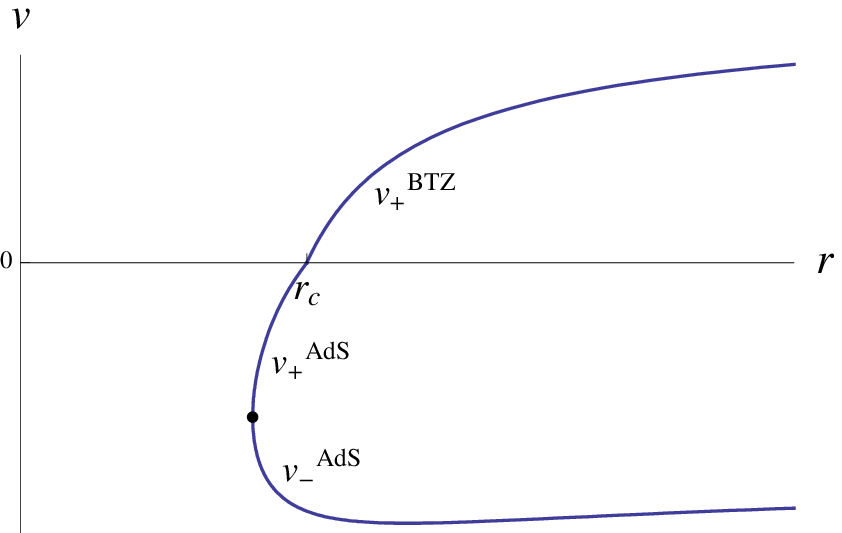}} & 
\subfigure[]{\includegraphics[width=6.8cm,height=3.3cm]{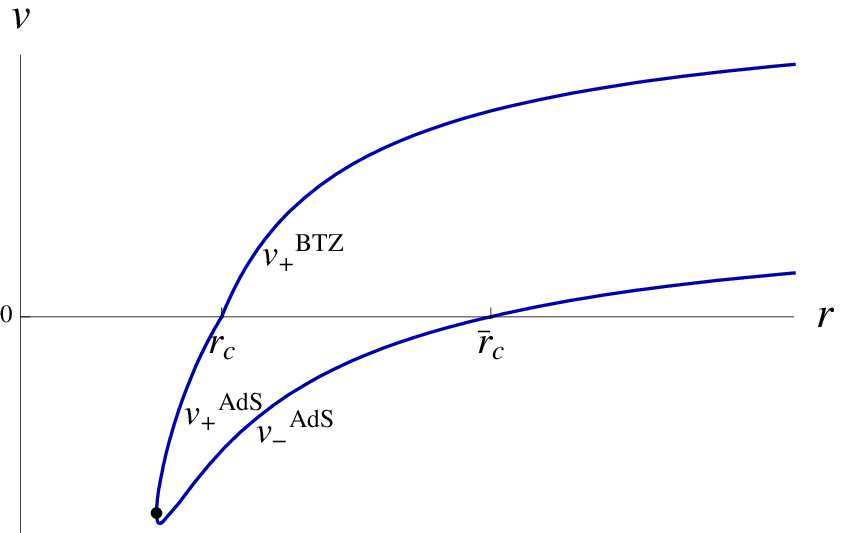}} \\
\subfigure[]{\includegraphics[width=6.8cm,height=3.3cm]{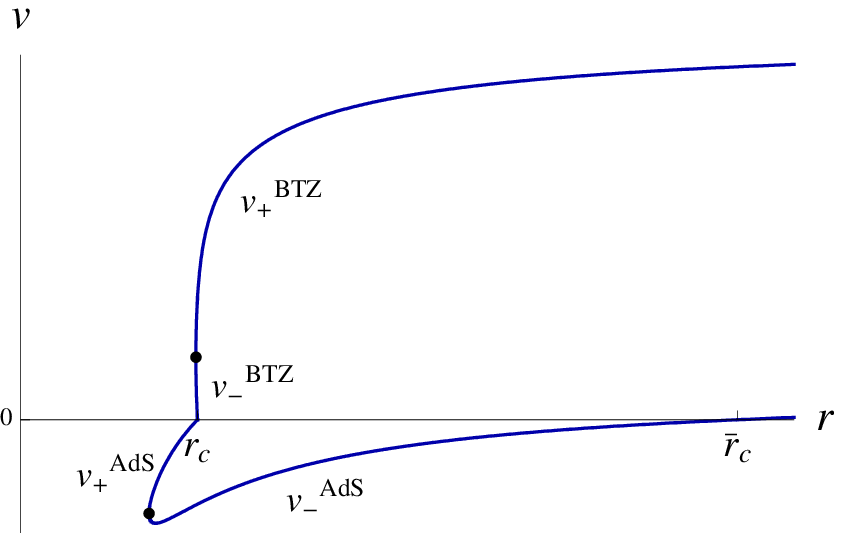}} & 
\subfigure[]{\includegraphics[width=6.8cm,height=3.3cm]{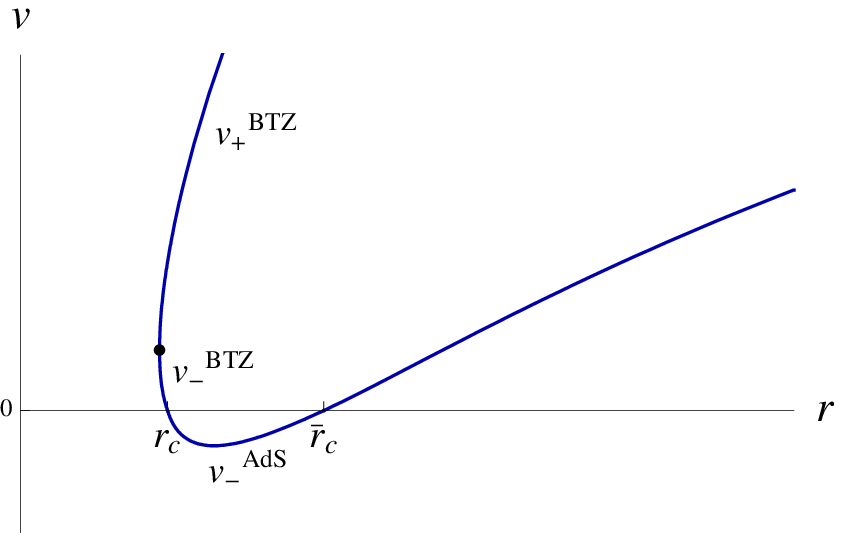}}
\end{tabular}
\caption{\label{fig:examplesdindout} Examples of different geodesics according to our classification in terms of $d_\text{in}$ and $d_\text{out}$. a) Geodesic with $t_1\!>\!0$, $t_2\!<\!0$ and $d_\text{in}\!=\!d_\text{out}\!=\!1$; Geodesics with $t_i\!>\!0$ and b) $d_\text{in}\!=\!d_\text{out}\!=\!1$; c) $d_\text{in}\!=\!1$ and $d_\text{out}\!=\!-1$; d) $d_\text{in}\!=\!d_\text{out}\!=\!-1$.}
\end{figure}

Using the matching conditions above, the construction of geodesics that intersect the shell can be done in a straightforward  although laborious way. We find it convenient to parameterize the geodesics in terms of $r_\ast$, $r_c\!=\!r(\lambda_c)$, $A_\text{in}$, which denotes the value of the energy constant for the geodesic segment inside the shell, and a sign $d_\text{in}$ describing in which branch the inner geodesic segment reaches the shell at $r_c$. The geodesics will intersect the shell at one or two points depending on whether $t_2$ is positive or negative, {\it i.e.} after or before the collapse of the shell starts. In case there are two intersection points, they will be in general located at different values of the radial coordinate. We choose $r_c$ to be the smaller one and we will call ${\bar r}_c$ the second, larger value in case it exits. As noticed in the previous subsection, when $A_\text{in}\!\neq\!0$ the minimal value of the radial coordinate reached by the inner geodesic segment, $r_m$ \eqref{rmsp}, and the point at which the minimal value of the null coordinate happens, $r_\ast$, do not coincide. Due to this, when $r_c\!<\!r_\ast$ the intersection point can happen in the branch $-$ of the inner segment, corresponding to $d_\text{in}\!=\!-1$, as shown in Fig \ref{fig:examplesdindout}d.
The previous data completely determines the shape of the geodesic, but not its position in the $x$ direction. Since the system is invariant under translations in $x$, this information does not affect any relevant physical quantity and we will ignore it.

The only restriction we will assume in the following is $r_\ast\!>\!A_{\rm in}$.  For AdS$_3$ this condition corresponds to geodesics with spacelike separated endpoints. It turns out that this choice is able to provide all geodesics with spacelike separated endpoints also in the collapsing shell geometry. For geodesics whose endpoints tend to be lightlike separated, this fact is easily proved. Indeed we have shown that in the lightlike limit geodesics remain tangent to the boundary, and close to the boundary the BTZ and AdS geometries coincide.  

From the matching condition \eqref{con3} we obtain the value of the energy parameter in the BTZ geodesic segment intersecting the shell at $r_c$  
\be
A _\text{out}={\big| A_\text{in}(2 r_c^2-m)-d_\text{in}  m \sqrt{A_\text{in}^2+r_c^2-r_\ast^2} \, \big| \over 2 r_c^2} \, .
\label{aout}
\ee
and the sign $d_\text{out}$ encoding its associated branch
\be
d_\text{out}=\epsilon \, \sign \! \left[  { m \sqrt{A_\text{in}^2+r_c^2-r_\ast^2}+ d_\text{in} A_\text{in} (m-2 r_c^2) \over (m-2 r_c^2) \sqrt{A_\text{in}^2+r_c^2-r_\ast^2} + d_\text{in} A_\text{in} m} 
\right] \, ,
\label{dout}
\ee
where $\epsilon$ is given by \eqref{ee}.
Contrary to the thermal case, when the entire geometry is described by a BTZ black hole, we need now to allow both positive and negative values for the constant $C$ \eqref{CC} in order to construct the outer geodesic segment.

\begin{figure}[thbp]
\centering
\begin{tabular}{cccc}
\hspace{-5mm}
\subfigure[]{\includegraphics[width=3.5cm,height=2.7cm]{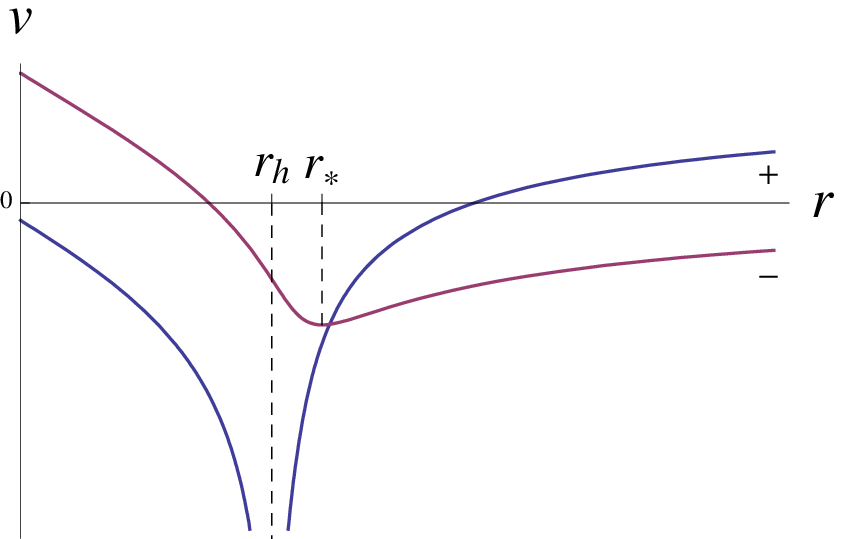}} &
\subfigure[]{\includegraphics[width=3.5cm,height=2.7cm]{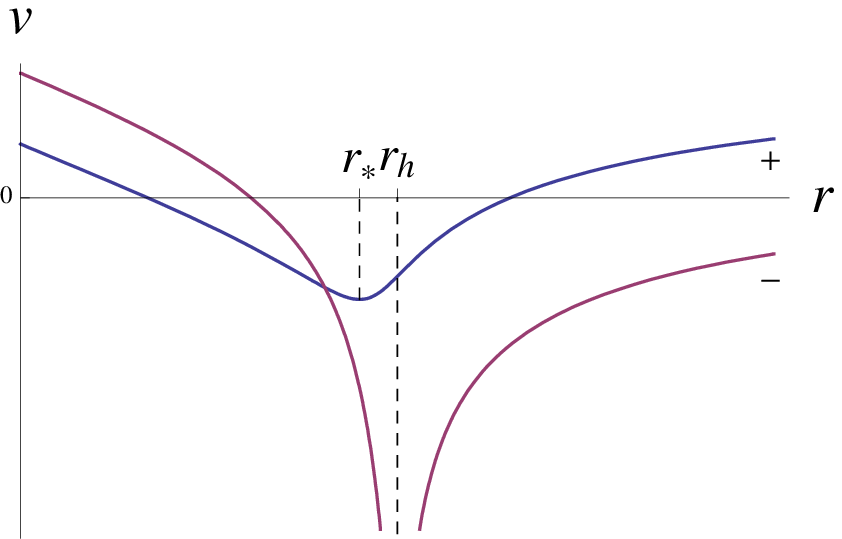}} &
\subfigure[]{\includegraphics[width=3.5cm,height=2.7cm]{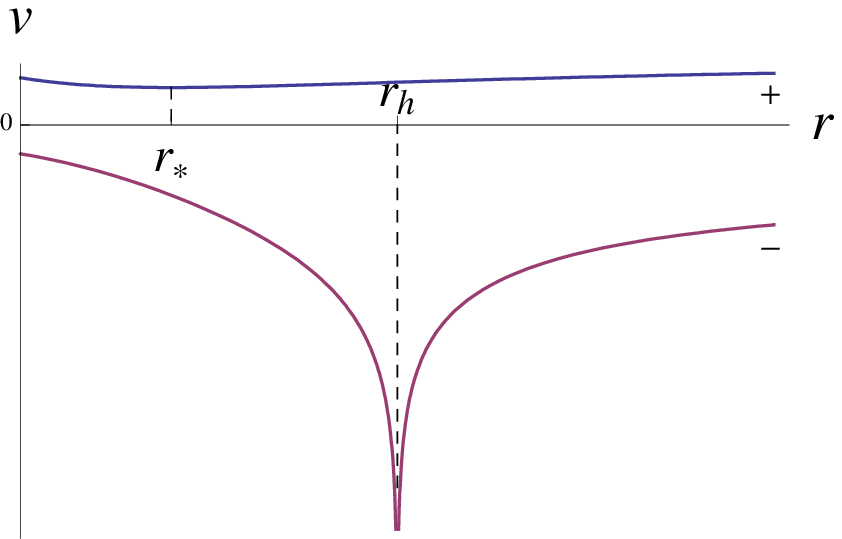}} &
\subfigure[]{\includegraphics[width=3.5cm,height=2.7cm]{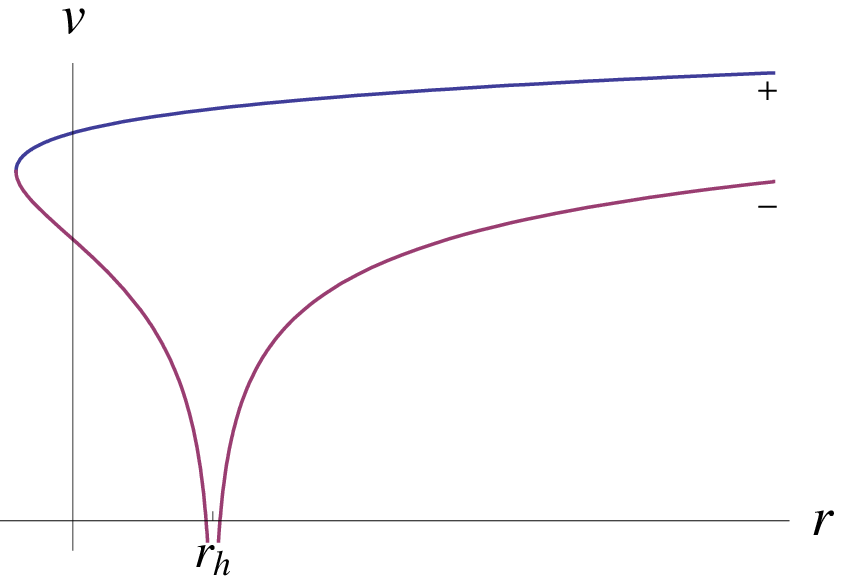}}
\end{tabular}
\caption{\label{fig:examples2BTZ} Examples of BTZ geodesics with a) $\epsilon\!=\!1$ and $C\!<\!0$; $\epsilon\!=\!-1$ and b) $C\!<\!0$; c) $C\!>\!0$ and ${\tilde \epsilon}\!=\!-1$; d) $C\!>\!0$ and ${\tilde \epsilon}\!=\!1$, in this case $r_\ast\!<\!r_m$.}
\end{figure}

Not all possible data for the inner geodesic lead to geodesic in the collapsing shell geometry that are relevant for the calculation of two-point functions. Indeed, some of them produce geodesics whose outer BTZ segment falls into the black hole singularity, instead of returning to the boundary. In Fig.\ref{fig:exampleBTZ} and Fig.\ref{fig:examples2BTZ} we have plotted all types of shapes that the (real part) of the functions $v_\pm(r)$ in the BTZ background can have\footnote{The position in the $v$ coordinate is not relevant in this case since it can be shifted without altering the shape of the functions due to the invariance under time translations.}. They are easily understood using the generalization of \eqref{dvdr} to an arbitrary geodesic
\be
{d v_\pm \over d r} ={1 \over r^2-m} \left( {\pm \, \epsilon A r \over \sqrt{A^2 r^2 +(r^2-m)(r^2-r_\ast^2)}} +1\right) \, .
\label{dvdr2}
\ee
When $\epsilon\!=\!1$ ($-1$) the branch $-$ ($+$) has a minimum at $r_\ast$, having a positive derivative for $r\!>\!r_\ast$ and negative otherwise. When $\epsilon\!=\!-1$ the branch $-$ always diverges to minus infinity at the horizon, having a positive derivative above it and negative below. When $\epsilon\!=\!1$ the same pattern for the derivatives holds in the branch $+$, but the horizon is only reached for $C\!<\!0$. An example of geodesic which should be discarded is in Fig.\ref{fig:examples2BTZ}a when $d_\text{out}\!=\!1$ and $r_c\!<\!r_h$, or when $d_\text{out}\!=\!-1$ and $r_c\!<\!r_\ast$. 

When $d_\text{out}\!=\!1$ the only way to avoid that the outer geodesic segment ends at the singularity is that $v'$ is positive at the merging point. If $d_\text{out}\!=\!-1$ and $v'$ at $r_c$ is negative, the outer geodesic segment must change into the branch $+$ before reaching the boundary. This can only happen in the situations exemplified in Fig.\ref{fig:exampleBTZ} and Fig.\ref{fig:examples2BTZ}d. Let us define an index $\eta$ which equals minus one if there is change of branches in the outer segment and one otherwise. We then have 
\be
d_{\rm out}=1\, \rightarrow \, \eta=1 \, , \hspace{1cm} d_{\rm out}=-1\, \rightarrow \, \left\{ \begin{array}{ll} \eta=\sign(r_c-r_\ast)  &\text{for} \; \epsilon=1 \\ \eta=\sign(r_c-r_h)  \;\;& \text{for} \;  \epsilon=-1 \end{array}
\right. \, .
\label{etaa}
\ee
For example, geodesics in Fig.\ref{fig:examplesdindout}c and \ref{fig:examplesdindout}d have $\eta\!=\!-1$. 
In Section 6 we will see what is the restriction on the data of the inner geodesic segment that ensures geodesics in the collapsing shell geometry not to fall into the singularity.

We will call $d_\text{in}^x$ and $d_\text{out}^x$ the signs describing the matching branches for the inner and outer $x(r)$ functions. The choice \eqref{bbcc} for the geodesic endpoints sets $d_\text{in}^x\!=\!d_\text{in}$. Combining the matching equations \eqref{con1} and \eqref{con2} we have
\be
d_\text{out}^x= \eta \, .
\label{signx}
\ee

Finally, a second intersection point of the geodesic with the shell will exist at 
\be
{\bar r}_c={ r_c \, r_\ast^2 \over r_\ast^2 -2 A_\text{in} ( A_\text{in}+d_\text{in} \sqrt{A_\text{in}^2+r_c^2-r_\ast^2}) } \, ,
\label{secondint}
\ee
whenever the denominator of this expression is positive \footnote{
For $r_\ast^2\!>\!2A_\text{in}^2$ this is always the case when $d_\text{in}\!=\!-1$, while it requires $r_0\!<\!r_\ast^2/2A_\text{in}$ for $d_\text{in}\!=\!1$. When instead $r_\ast^2\!<\!2A_\text{in}^2$, the denominator is positive for $r_0\!>\!r_\ast^2/2A_\text{in}$ if $d_\text{in}\!=\!-1$, while it is always negative for $d_\text{in}\!=\!1$.}. For those cases where there are two intersection points with the infalling shell, the two separate outer geodesic segments will have in general different energy parameters $A_\text{out}$ and ${\bar A}_\text{out}$. Also $d_\text{out}$ and ${\bar d_\text{out}}$ might differ. The matching conditions \eqref{con1} and \eqref{con2} for the second intersection point lead to
\be
{\bar d}_\text{out}^x=- {\bar \eta}  \, .
\label{signx}
\ee

We have now all the data necessary to construct the complete geodesic in terms of the data of its inner segment. The explicit expression can be found in Appendix A. 
The proper length of the geodesic is $\int d\lambda\!=\!\int dr/{\dot r}$, substituting \eqref{B} with the constants appropriate for each segment. The length of the inner segment is given by
\be
L_\text{in}= d_\text{in} \log \big[ r_m^{-1} \big(r_c+ \sqrt{r_c^2-r_m^2} \big)\big] + \log \big[ r_m^{-1} \big(r_M+ \sqrt{r_M^2-r_m^2} \big)\big]  \, .
\label{geolenads}
\ee
When the geodesics intersects twice the infalling shell, $r_M\!=\!{\bar r}_c$. Otherwise $r_M$ should be replaced by a large but finite value $r_b$, as explained in previous sections. Inverse powers of $r_b$ are then to be neglected in the second term of the rhs. The length of the outer geodesic segment that starts at $r_c$ is 
\be
L_\text{out}  =  \log 2 \, r_b + {1 \over 2} \log \left| C^{-1} \big( r_\ast^2 - A_\text{out}^2 +m-2 r_c^2+2 \eta \sqrt{r_c^4+m r_\ast^2-r_c^2 ( r_\ast^2 - A_\text{out}^2 +m)} \big)  \right| \, . 
\label{geolenbtz}
\ee
In the case that the geodesic intersects twice with the shell, there is an analogous expression for the length ${\bar L}_\text{out}$ of the outer geodesic segment starting at ${\bar r}_c$.

\begin{figure}[thbp]
\centering
\begin{tabular}{cc}
\subfigure[]{\includegraphics[width=7cm]{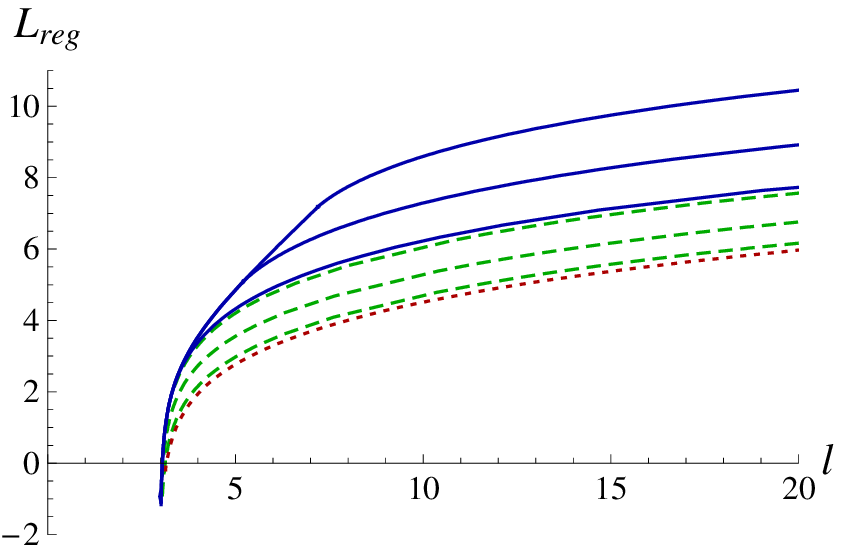}} & 
\subfigure[]{\includegraphics[width=7cm]{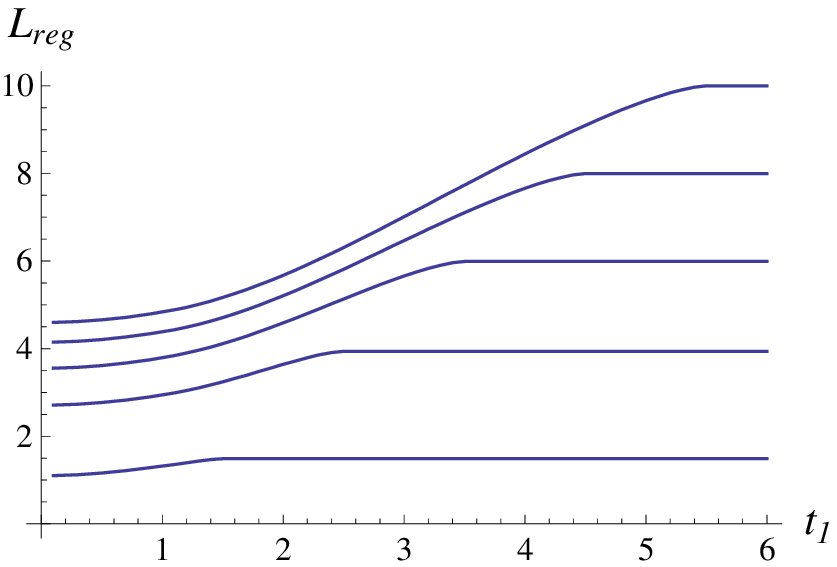}}
\end{tabular}
\caption{\label{fig:fixedTDT} For $m\!=\!1$ regularized geodesic length a) as a function of $l$ for fixed $\Delta t\!=\!3$ and $t_1\!=\!3.1,4.1,5.1$ from bottom up in solid lines ($t_2>0$), $t_1\!=\!0.9,1.9,2.9$ in dashed lines ($t_2<0$) and empty AdS geodesic in dotted line; b) as a function of $t_1$ for fixed $\Delta t\!=\!1$ and $l\!=\!2,4,6,8,10$ from  bottom up respectively.}
\end{figure}

In Fig.\ref{fig:fixedTDT}a we have plotted the regularized geodesic proper length \eqref{lreg} as a function of $l$ for fixed $\Delta t\!=\!t_1\!-\!t_2$. It diverges when the geodesics approach the lightlike limit. In dashed lines we have distinguished geodesics with negative $t_2$, those associated to two-point functions with one time after and one before the perturbation acts. When $t_1\!<\!0$ the result is that of empty AdS$_3$. In solid lines we have plotted the length of geodesic with both endpoints after the perturbation.   

As we have shown in subsection 4.1, geodesics with both $t_i$ positive and $l\!<\!t_1\!+\!t_2$ lie completely outside the shell. Perceiving a BTZ geometry they reproduce the results at thermal equilibrium. Namely for $l$ bigger than $\Delta t$ and such that the associated geodesics can reach the black hole horizon\footnote{This distance is of order $\beta/2$ for equal-time geodesics.}, the geodesic length grows linearly 
\be
L_{\text{reg}}(l;t_1,t_2)=2\log {\beta \over 2 \pi}+{2 \pi l \over \beta} \, .
\label{extensive}
\ee
A linear regime for $L_\text{reg}$  can be clearly seen in Fig \ref{fig:fixedTDT}a.
Geodesics with $l\!>\!t_1\!+\!t_2$ intersect the shell and lead to a result that deviates from thermal. 
For asymptotically large intervals the geodesic length returns to the same logarithmic behavior as in AdS$_3$, plus a time-dependent shift
\be
L_{\text{reg}}(l;t_1,t_2)=2\log l +s(t_1,t_2) \, .
\label{loggrowth}
\ee
Plots of the geodesic length as a function of $t_1$ for fixed $l$ and $\Delta t$ are shown in Fig.\ref{fig:fixedTDT}b. It saturates to its thermal value at $t_1\!=(l+\Delta t)/2$, when the associated geodesic stops reaching the infalling shell. 

Although our results are exact, $L_{reg}$ is a complicated function of the geodesic data which can not be explicitly expressed in terms of the physical variables $t_1$, $t_2$ and $l$. Explicit results can be however derived for large intervals or late times. The next sections are devoted to analyze the evolution of two-point functions in these two limits, from which we will derive important physical information.

\section{Two-point functions for asymptotically large separations}

When $l$ is very large, the associated geodesic will reach deeply inside the shell. In particular, for $l \!\rightarrow\!\infty$ the geodesic will get closer and closer to $r\!=\!0$. We want to take the large $l$ limit while keeping $t_1$ and $t_2$ finite. Since the major part of the geodesic will be contained inside the shell, in the AdS part of the geometry, relation  
\eqref{tlAdS} implies 
\be
r_\ast \simeq {2 \over l} \, , \;\;\;\;\; A_\text{in}\simeq{2a \over l^2} \, , 
\label{rl}
\ee
for some finite constant $a$, and necessarily $d_\text{in}\!=\!1$. The parameters of the outer piece of the geodesic are obtained from \eqref{aout} and \eqref{dout}
\be
A_\text{out}={m \over 2 r_c} \, , \;\;\;\;\; d_\text{out}=\sign  (2 r_c^2-m) \, .
\label{longpar}
\ee
If the first crossing happens at the value of the radial coordinate $r_c$, the second should take place at ${\bar r}_c\!=\!r_c/(1-a r_c )$. Hence there will be a second intersection point only if $a\!<\!1/r_c$, in which case expressions analogous to \eqref{longpar} hold.

The length of the AdS geodesic segment,  given by \eqref{geolenads}, is 
\be
L_\text{in}= 2\log l +\log r_c +\log r_M +{\cal O}(l^{-1})\, ,
\label{ladsinf}
\ee    
where as before $r_M$ equals ${\bar r}_c$ if the geodesic intersects twice with the shell and $r_b$ otherwise.
Here $l$ strictly represents the space separation between the endpoints of the inner geodesic, prolonged to the boundary in the absence of the shell. However the difference between this separation and that calculated in the presence of the shell, having into account the matching conditions, is a finite number. Thus \eqref{ladsinf} applies equally with $l$ being the space separation between the geodesic endpoints in the thin shell geometry. The length \eqref{geolenbtz} of the outer geodesic segment starting at $r_c$ and its endpoint time $t_1$, given by \eqref{t1}, are 
\be
L_\text{out}=\log 2 \,r_b -\log{4r_c^2-m \over 2 r_c}  \, , \;\;\;\; t_1={1 \over r_h} \log {2r_c+r_h \over 2r_c-r_h} \, ,
\label{trc2}
\ee
with analogous expressions holding for the second crossing at ${\bar r}_c$. We have used the fact that $r_c\!>\!r_h/2$ for $r_\ast\!\simeq\!0$, which will be explained in the next section. The relation between $t_i$ and the associated intersection point can be easily inverted. Adding the length of the several geodesic segments and subtracting $2 \log r_b$, we finally obtain 
\be
L_{\text{reg}} =2\log l +{\tilde s}(t_1) + {\tilde s}(t_2) +{\cal O}(l^{-1})\, ,
\label{lreglarge}
\ee
where
\be
{\tilde s}(t) = \; \left\{ \begin{array}{ll} 2 \log [ \cosh ( \pi t /\beta)]  \, , &\;\;\;\;\; t>0 \\ 0 \, , & \;\;\;\;\; t<0 \end{array} \right. \, .
\label{s}
\ee
Comparing with \eqref{loggrowth}, we have $s(t_1,t_2)\!=\!{\tilde s}(t_1) + {\tilde s}(t_2)$. This precisely matches the numerical results of \cite{AbajoArrastia:2010yt}.

Substituting \eqref{s}, we obtain the asymptotic value of the two-point function
\be
\langle {\cal O}(l,t_1) {\cal O}(0,t_2) \rangle\simeq  {1 \over \big(l \cosh(\pi t_1/\beta) \cosh(\pi t_2 / \beta) \big)^{2\Delta}} \, .
\label{2plargel}
\ee
Its dependence on $l$ reproduces that of the two-point function on the CFT vacuum for large spatial distances. On the contrary, an exponential suppression with $l$ was found in the same regime after a quantum quench, see \eqref{noneq2p} \cite{Calabrese:2006rx,Calabrese:2007rg}. Both quantum quenches and our holographic model generate at $t\!=\!0$ a sea of excitations with non-trivial entanglement properties. From then on coherent excitations start separating from each other.
However if we consider very large $l$, for any finite $t_i$ they will have not separated enough to erase the effect of the initial entanglement pattern on the two-point functions. Hence the different behavior of the two-point functions in both dynamical situations should be traced back to the different initial states that trigger the evolution. In the holographic set up, the perturbation modeled by the shell does not destroy the long range correlations proper of the CFT vacuum \cite{AbajoArrastia:2010yt}. However the ground state of a massive hamiltonian was taken as initial state for the quenches considered in \cite{Calabrese:2006rx,Calabrese:2007rg}. In an state with a mass gap, the exponential damping of correlators with the distance is expected. 

Another important feature of \eqref{2plargel} is its increasing suppression as a function of time due to the contribution of the ${\tilde s}(t_i)$. This function has a linear behavior for times $t\!\gtrsim \! \beta$
\be
{\tilde s}(t) \simeq -2\log 2 +{2 \pi t \over \beta} \, ,
\label{slinear}
\ee 
as noted in \cite{AbajoArrastia:2010yt}. Let us denote $\bar t$ the threshold time from which on the previous approximation is valid. The time dependence of the two-point function \eqref{2plargel} for $t_i\!>\!{\bar t}$ reduces to $e^{-2 \pi \Delta (t_1\!+\!t_2)/\beta}$. Identifying $\beta\!=\!4\tau_0$, this is in agreement with the quantum quench result.

\section{\label{sec:latet}Steady evolution}

For finite separations $l\!>\!t_1\!+\!t_2$, the $l$ dependence of two-point functions after a quantum quench turns out to enter in terms of the effective distance $l\!-\!t_1\!-\!t_2$, as seen in \eqref{noneq2p}. 
We will show in this section that the same pattern applies to the holographic model for times after the threshold $\bar t$. Namely, the power law behavior of two-point functions with $l$ that we found in the previous section for very large distances, gets promoted for finite $l\!>\!t_1\!+\!t_2$ and $t_i\!>\!{\bar t}$ to a function of $l\!-\!t_1\!-\!t_2$.
The evolution of the entanglement entropy in the same holographic set up verifies an analogous property. It was shown numerically in \cite{AbajoArrastia:2010yt} that the entanglement entropy of a single interval of size $l$, for $l\!>\!2t$ and $t$ after the same threshold time ${\bar t}$, is a function of the combination  $l\!-\!2t$.

In order to arrive to the above conclusion, we first need to recall some further properties of the geodesics. As mentioned in Section 4, not all the initial data $r_\ast$, $r_c$, $A_\text{in}$ and $d_\text{in}$ lead to geodesics relevant for the computation of two-point functions. Indeed some of them produce geodesics that fall into the black hole singularity instead of returning to the boundary. The set of initial data that avoid this problem is 
\be
\begin{array}{ll}
r_\ast>r_h+A_\text{in} \; : & \left\{ \begin{array}{ll} r_c\in [r_m,\infty] \, , & d_\text{in}=1 \\  r_c\in [r_m,r_\ast] \, , \;\;\;\;\;  & d_\text{in}=-1 \end{array} \right. \, , \\[5mm]
r_\ast\in [{\tilde r}_\ast,r_h+A_\text{in}] \; : & \left\{ \begin{array}{ll} r_c\in [r_m,\infty]  \, , & d_\text{in}=1 \\  r_c\in [r_m,R]  \, , \;\;\;\;\; & d_\text{in}=-1 \end{array} \right. \, , \\[5mm]
r_\ast\in (A_\text{in},{\tilde r}_\ast] \; : & \;\;\;\; r_c\in[R,\infty] \, , \;\;\;\;\;\; d_\text{in}=1 \, .
\end{array}
\label{rangeint}
\ee
where 
\be
R=r_h \, {r_h + 2 A_\text{in} - 2 r_\ast + \sqrt{(r_h + 2 A_\text{in} )^2 + 4 r_\ast (r_h-r_\ast )}) \over 4 (r_h + A_\text{in}-r_\ast)}  \, .
\label{r0ex}
\ee
and ${\tilde r}_\ast$ is the only value of $r_\ast$ for fixed $A_\text{in}$ at which $R\!=\!r_m$, with $r_m$ the minimal radius of spacelike AdS$_3$ geodesics \eqref{rmsp}. The functional dependence of ${\tilde r}_\ast$ on $A_\text{in}$ is given in \eqref{rslimit}. For any other $r_\ast$, $R\!>\!r_m$. Consistently at $r_c\!=\!r_h\!+\!A_\text{in}$, $R\!=\!r_\ast$.  See Appendix B for a discussion of the properties of the functions $R$ and ${\tilde r}_\ast$.

We have derived the bound $R$ on the intersection point by inspection over the parameter space of the geodesics. The casuistic for the construction of generic geodesics is so broad that unfortunately we do not have a closed proof for \eqref{r0ex}. 
An illustrative example for the necessity of introducing $R$ is provided by $r_\ast \!=\!A_\text{in}\!=\!0$. From \eqref{longpar} we know that when $r_c\!<\!r_h/\sqrt{2}$ the inner geodesic merges at the shell with the branch $-$ of a BTZ geodesic. Using \eqref{etaa} we have that for these values $\eta\!=\!-1$. For $r_c\!=\!r_h/2$, the minimal radial value \eqref{rmin} at which the branches $\pm$ of the BTZ geodesic join happens at the singularity. For smaller values of $r_c$, ${\tilde \epsilon}$ becomes negative implying the outer geodesic will have two disconnected pieces each of them reaching to the singularity. Hence $r_c$ must be bigger than $R\!=\!r_h/2$ in order to have an smooth outer geodesic that returns to the boundary. 

A very important property of \eqref{r0ex} is that as $r_c, {\bar r}_c\!\rightarrow\! R$, the functions $t_i$, $l$ and $L_{reg}$ turn out to diverge. Let us define $\delta r_{c}\!=\!|r_{c}\!-\!R|/R$ and $\delta {\bar r}_{c}\!=\!|{\bar r}_{c}\!-\!R|/R$. In order that both $r_c$ and ${\bar r}_c$ tend to $R$, definition \eqref{secondint} implies that $A_\text{in}$ must be small. In particular, $A_\text{in}$ must be of the order of $\delta r_{c}$ and $\delta {\bar r}_{c}$. We show in Appendix B that in this limit
\be
\begin{array}{l}
t_1=  \displaystyle -{1 \over 2 r_h} \log \delta r_c +{1 \over 2 r_h} \log \left| {2 r_h (r_h-r_\ast)(R_0^2-r_h r_\ast)  \over r_\ast f_0 (r_h-R_0)^2} \right|  \, , \\[7mm]
t_2 =  \displaystyle -{1 \over 2 r_h}  \log \delta {\bar r}_c +{1 \over 2 r_h} \log \left| {2 r_h (r_h-r_\ast)(R_0^2-r_h r_\ast)  \over r_\ast f_0 (r_h-R_0)^2} \right|  \, , \\[7mm]
l =  \displaystyle -{1 \over 2 r_h} \log (\delta r_{c} \delta {\bar r}_{c} )  + {1 \over 2 r_h} \log \left| {2 r_h r_\ast (R_0^2-r_h r_\ast)  \over (r_h-r_\ast) f_0 R_0^2} \right|+ {4 (r_h-r_\ast) R_0 \over m r_\ast}  \, , \\[7mm]
L_{reg} =  \displaystyle  -{1 \over 2} \log (\delta r_{c} \delta {\bar r}_{c} )  + \log \left| {2 (R_0^2-r_h r_\ast) \over f_0 r_h r_\ast (r_h-r_\ast)} \right| + 2\log {R_0 + \sqrt{R_0^2-r_\ast^2} \over r_\ast} \, .
\end{array}
\label{Rlimit}
\ee
where $R_0$ is the value of $R$ \eqref{r0ex} for $A_\text{in}\!=\!0$, and $f_0$ a function of $r_\ast$ whose expression will not be relevant for us. These relations are valid up to terms of order $\delta r_{c}$, $\delta {\bar r}_{c}$ and $A_\text{in}$.

Combining the previous equalities, we obtain 
\be
l-t_1-t_2={2 \over r_h} \log {(r_h-R_0) r_\ast \over (r_h-r_\ast) R_0 } + {4 (r_h-r_\ast) R_0 \over m r_\ast}  \, .
\label{xr}
\ee
The importance lesson here is that the rhs is only function of $r_\ast$. From \eqref{rangeint} we know that $R$ only plays a role for $r_\ast\!\leq\!r_h\!+\!A_\text{in}$. Since we are taking the inner energy parameter to be zero, we must consider $r_\ast$ in the interval $[0,r_h]$. Consistently, this range covers univocally values of the lhs from just the threshold for thermal behaviour at $r_\ast\!=\!r_h$ where we have
\be
l-t_1-t_2=0 \, ,
\ee
to infinite spatial separation for the geodesic endpoints as $r_\ast \!\rightarrow\! 0$, when \eqref{xr} leads to
\be
l\simeq {2 \over r_\ast} \, ,
\ee
as previously derived in \eqref{rl}.

\begin{figure}[thbp]
\centering
\begin{tabular}{cc}
\subfigure[]{\includegraphics[width=7.8cm]{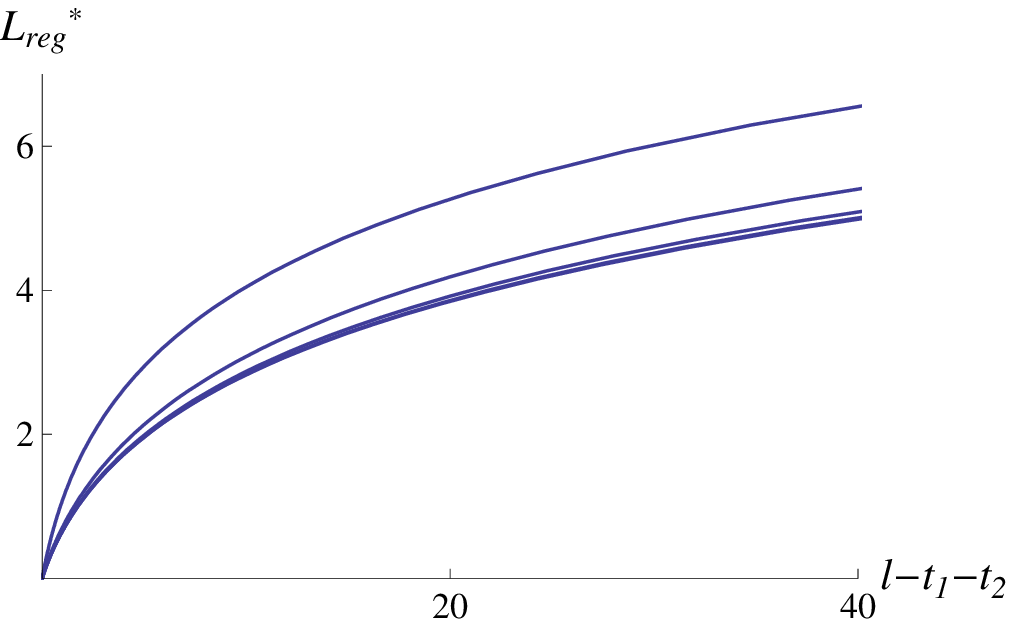}} & 
\subfigure[]{\includegraphics[width=6.8cm]{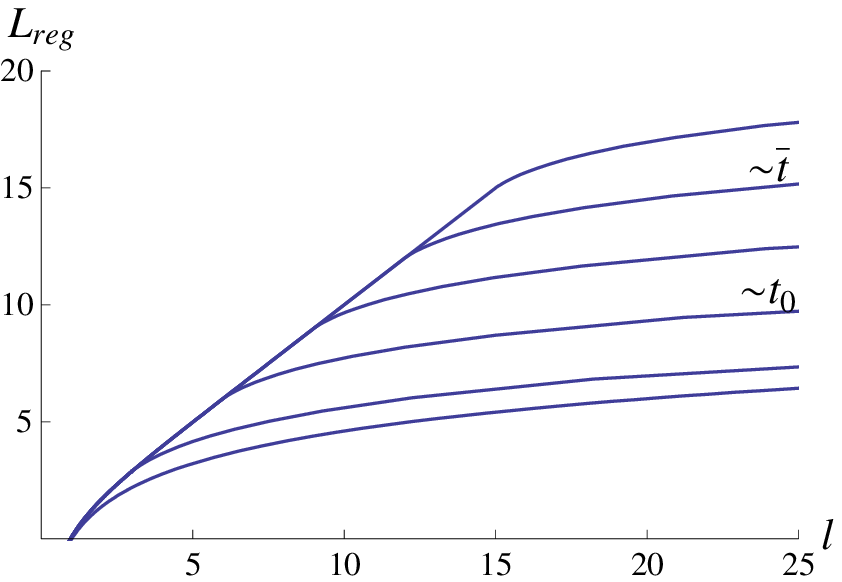}} 
\end{tabular}
\caption{\label{fig:latet} For $m\!=\!1$, a) $L_\text{reg}^\ast \!= \!L_\text{reg}\!-\!L_\text{reg}(l\!=\!t_1\!+\!t_2)$ as a function of $l\!-\!t_1\!-\!t_2$ for $t_1\!=\!1.5,2.5,3.5,4.5,5.5$ from top down and $\Delta t\!=\!1$. The last two lines practically coincide with the function $F$ in \eqref{steady} (up to a constant). b) $L_\text{reg}$ as a function of $l$ for $\Delta t\!=\!0$ and $t\!=\!0,1.5,3,4.5,6,7.5$. We have signaled approximately the times $t_0$ and $\bar t$.}
\end{figure}

Using \eqref{Rlimit}, we obtain for the geodesic length
\be
L_{\text{reg}}= r_h (t_1+t_2) -2\log r_h + 2\log { r_h-R_0 \over r_h-r_\ast }+2 \log { R_0+\sqrt{R_0^2-r_\ast^2} \over r_\ast} \, .
\label{Lr}
\ee 
At $r_\ast\!=\!r_h$, the threshold for thermal behavior, this expression reproduces the extensive dependence \eqref{extensive} characteristic of thermal equilibrium. At $r_\ast \!\rightarrow\! 0$ we have
\be
L_{reg}=r_h(t_1+t_2)-4 \log 2 +2 \log l \, .
\ee 
Agreement with the result \eqref{lreglarge} for the geodesic length at large space separations requires $t_i\!>\!{\bar t}$. Extracting the values of $t_i$ for which \eqref{Rlimit} applies is involved. However using $r_h\!=\!2\pi /\beta$, a rough analysis indicates that they are consistent with the bound set by ${\bar t}$.
Hence, inverting \eqref{xr} to obtain $r_\ast$ as a function of $l\!-\!t_1\!-\!t_2$, we derive the announced result
\be
L_{\text{reg}}={2 \pi \over \beta}(t_1+t_2)+F(l-t_1-t_2)  \, ,
\label{steady}
\ee
for $t_i\!>\!{\bar t}$ and separations $l\!>\!t_1\!+\!t_2$. 
The validity of the above expression can be further checked in Fig. \ref{fig:latet}a, where we have plotted the geodesic length as a function of $l\!-\!t_1\!-\!t_2$ for several values of $t_1$ keeping $\Delta t$ fixed. We observe that as $t_1$ increases the curves tend to coincide up to a shift, as implied by \eqref{steady}. 

We can identify two sources that drive the evolution of an out of equilibrium system towards equilibration: the interaction between its components and their propagation. In the sea of excitations sourced by the initial perturbation, quantum correlations will be stronger between excitations generated at close points. Propagation will tend to increase the separation between entangled excitations with time. On the other hand interaction will lead to the redistribution of energy, and other conserved charges in case they are present, among the different momentum modes. This is crucial in order that the state towards which the system evolves is that of thermal equilibrium. The separation of entangled excitations with time is a process constrained by causality, implying that equilibration can only be achieved on regions of growing but finite size. On the contrary, the redistribution of energy among modes can be expected to require a finite time for its nearly completion.

In Section 2 we have reviewed the simple model proposed in \cite{Calabrese:2005in} to account for the evolution pattern of the entanglement entropy and two-point functions after a quantum quench. It explained the dependence of two-point functions on the effective distance $l\!-\!t_1\!-\!t_2$ for separations $l\!>\!t_1\!+\!t_2$. The main hypothesis of this model was that the excitations generated by the perturbation move as free particles according to their group velocities, which for a 2-dimensional CFT is always given by the speed of light. Therefore it seems consistent to assume that when this model is applicable, all processes in the evolving plasma related with interaction have reached equilibrium. In this sense, notice that the general results for the evolution after a quantum quench derived in \cite{Calabrese:2005in,Calabrese:2006rx,Calabrese:2007rg} are only valid for times much larger than the inverse effective temperature. As it was proposed in \cite{AbajoArrastia:2010yt} based in the study of the entanglement entropy, this suggests to interpret ${\bar t}$ in our holographic model as the thermalization time for occupation numbers. Here we have confirmed the same conclusion from the evolution of general two-point functions.

It is interesting to cross check our arguments with the simple example of a quench from a massive to massless free boson, studied in \cite{Calabrese:2007rg}. In this case the occupation numbers are conserved along the evolution. They do not coincide with a thermal distribution, except for momenta smaller than the effective temperature $T_{eff}\!=\!m/4$, where $m$ is the mass before the quench. This provides a model where equilibration is not synonymous of thermalization. Since the occupation numbers are conserved we should have ${\bar t}_{free}\!=\!0$. An exact treatment is possible for all times and distances, and the following relation turns out to be fulfilled
\be
\langle {\cal O}(t,l)  {\cal O}(t,0) \rangle  = e^{-\Delta(2{\tilde s}(t)+ F(l-2t) )}
\ee
for any time after the quench and distances $l\!\gtrsim\!\beta_{eff}$\footnote{For the free boson quench we have
\be
{\tilde s}(t)=-{1 \over 2} f(2mt) \, , \hspace{1cm} F(x)={\pi \over 4} mx -{1 \over 2} + {1 \over 2} f(mx) \, , 
\label{freeq}
\ee 
where $f(x)\!=\!1\!+\!{1 \over 2} G^{21}_{13}\Big( {x^2 \over  4} \Big| \begin{array}{ccc}  & \!\!{\scriptstyle 3/2}\!\! & \\[-1mm] \! {\scriptstyle 0} & \!\!{\scriptstyle 1}\!\! & {\scriptstyle 1/2} \! \end{array} \Big)$ with $G$ the Meijer G-function.}. Notice that also \eqref{steady} requires a restriction on the distances where it applies. The function ${\tilde s}$ for the free quench \eqref{freeq} has a similar structure to that of the holographic model. It vanishes at $t\!=\!0$ and after some time it enters a regime of linear growth. The main difference with the holographic model is that  the time at which $\tilde s$ behaves linearly does not coincide with ${\bar t}_{free}$. This example suggests that the important property which could identify the equilibration of occupation numbers in the models we are considering, is when the dependence on the scale $l$ enters through the effective distance $l\!-\!2t$ or $l\!-\!t_1\!-\!t_2$.  

In the free boson model the time at which ${\tilde s}$ starts growing linearly coincides with that at which $L_{reg}$ begins to show an extensive behavior for some range of $l$. Let us call the time defined by this second property $t_0$. In the holographic model $t_0\!\sim\!\beta/4$. It is the time at which equal-time geodesics anchored on the boundary start reaching the black hole horizon and produce a result clearly identifiable as thermal. It can be associated through propagation at the speed of light with the smallest scales able to feel the thermal bath. The time ${\bar t}$ is clearly bigger than $t_0$ in the holographic model (see Fig.\ref{fig:latet}b). The fact that the latter can be explained from propagation provides an additional argument in favor to relate the former with interaction.

\section{Discussion}

In this paper we have studied the evolution of two-point functions in a 2-dimensional CFT using holographic techniques. In our model the CFT starts in its vacuum state and at $t\!=\!0$ undergoes an instantaneous perturbation that brings the system out of equilibrium. The dual gravitational background that represents this process is that of the collapse of an infinitely thin shell of null dust in a spacetime with negative cosmological constant. It has been argued in \cite{AbajoArrastia:2010yt,Takayanagi:2010wp} that this set up represent a unitary process in the field theory, implying that information is not lost at the microscopical level. However very general arguments \cite{reimann} show that the system will evolve towards an state that at the macroscopical level can be described as thermal equilibrium. This is reflected holographically in the formation of a black hole out of the gravitational collapse of the null dust shell.

An important question is how is the energy is distributed among modes of the initial excitations created by the perturbation. Both the Vaidya metrics and their thin shell limit asymptote at large radius to AdS$_3$, differing from it only for values of the radial coordinate set by the mass function of the black hole being formed. Since this is the only scale in the problem, it suggests that excitations are produced around energies of order $\sqrt{m(t)}$ at a rate of the order $d\sqrt{m(t)}/dt$. Hence UV excitations would be much suppressed in the initial state \cite{AbajoArrastia:2010yt}\footnote{This is also the case in the free boson quench \cite{Calabrese:2007rg}}. This is supported by the fact that both entanglement entropy and two-point functions never deviate from their vacuum results in the UV limit. Recently a holographic set up which allows flexibility on the initial conditions for the evolution has been proposed \cite{Heller:2011ju}. It would be very interesting to study the dynamics of non-local observables in these cases.

Although the previous reasoning indicates that the initial excitations are probably not produced very far from their equilibrium distribution in our model, a deviation from it is of course to be expected. Based on the holographic analysis, a time $\bar t$ after which the energy redistribution among modes according to thermal equilibrium should be nearly completed can be proposed \cite{AbajoArrastia:2010yt}. One of the main result of this paper is that the same time $\bar t$ derived from the analysis of the entanglement entropy, applies to generic two-point functions. After $\bar t$ and for $l\!>\!t_1\!+\!t_2$, the dependence on the spatial separation of two-point functions enters only through the effective combination $l\!-\!t_1\!-\!t_2$ in accordance with the late time evolution after a quantum quench \cite{Calabrese:2006rx,Calabrese:2007rg}. We have argued that this is the relevant property to identify the equilibration time for occupation numbers in these 2-dimensional systems. 

Both in quantum quenches and in our holographic model the initial sea of excitations has a non-trivial entanglement pattern, where the main difference between both cases is the presence of long range entanglement in the latter. In order that observables over an scale $l$ can reproduce a thermal result, it is necessary that entangled excitations have separated at least a distance $l$. Causality then implies that thermalization can not be achieved globally over the system. For 2-dimensional CFT's, all excitations propagate at the speed of light. This gives rise to a sharp cut between the size of regions over which thermalization is effective and those in which it is not. A second central result of this paper is to reproduce the condition for thermal behavior of general two-point functions, $l\!<\!t_1\!+\!t_2$ with $t_i\!>\!0$, first derived in the study of quantum quenches \cite{Calabrese:2006rx,Calabrese:2007rg}.

When the dynamics of an out of equilibrium system is unitary, a very interesting question is how to define a coarse grained entropy that gives a good description of its evolution towards macroscopical equilibration.
In \cite{Takayanagi:2010wp} it was proposed that an interesting definition of a coarse grained entropy for compact systems could be given in terms of the entanglement entropy of a region $A$, where $A$ covers half of the total system, through
\be
S_{\text{eff}}(t)=2\big(S_A(t)-S_A^0\big) \, ,
\label{coarse}
\ee
with $S_A^0$ the entanglement entropy in the state before the evolution starts. A natural extension of this relation to the non-compact set up we are considering is to substitute the region $A$ by the complete non-compact system. Both for the 2-dimensional quantum quenches studied in \cite{Calabrese:2005in} and the holographic model based on the infalling shell of null dust \cite{AbajoArrastia:2010yt}, the entanglement entropy of a single asymptotically large interval satisfied
\be
S_\infty(t)={c \over 3} {\tilde s}(t) +S^0_\infty \, .
\ee
In the infinitely thin shell limit of the holographic model ${\tilde s}(t)$ is given by \eqref{s}. For the critical quantum quenches this function is only known for late times, where it coincides with \eqref{slinear} up to possibly a numerical factor. Therefore the generalization of \eqref{coarse} to the non-compact case would be 
\be
S_{\text{eff}}(t)={c \over 3} {\tilde s}(t) \, .
\label{coarseNC}
\ee
This definition implies that for any finite time $S_{\text{eff}}$ is finite. Hence no finite entropy density is generated, in agreement with the fact that equilibration is only effective over finite regions. Notice that this rules out the apparent horizon, in the theories with a holographic dual, as the place where the field theory entropy resides. Moreover in the thin shell model the apparent horizon, which is located at $r\!=\!\sqrt{m} \, \Theta(v) $, would predict an instantaneous thermalization for the entropy. On the contrary the function $\tilde s$ shows an smooth interpolation between zero and the thermal density multiplied by the size of the largest thermalized region for late times. 

As we mentioned in Section 6, the function $\tilde s$ for the free boson quench \eqref{freeq} has a similar structure to that of the holographic model, which describes thermalization in a strongly coupled plasma. The main difference between them affects the time at which they start to behave linearly. It would be very interesting to better understand the information about an equilibration process that can be extracted from \eqref{coarse} and \eqref{coarseNC}.

\begin{figure}[thbp]
\centering
\begin{tabular}{cc}
\includegraphics[width=7.2cm]{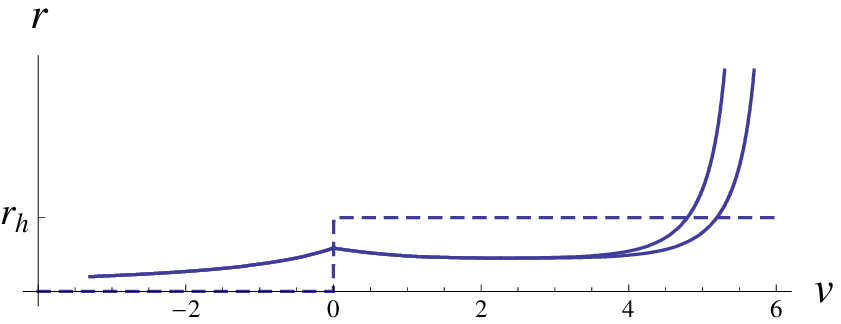} &\includegraphics[width=7.2cm]{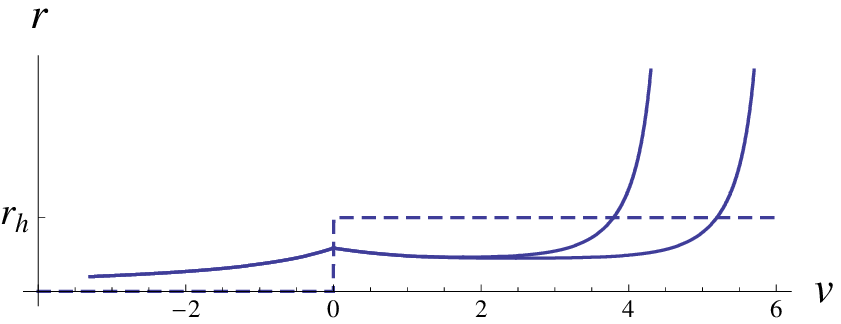}
\end{tabular}
\caption{\label{fig:behindhorizon} Geodesics whose outer segments are of the type that would connect the two boundaries of the extended BTZ Penrose diagram. They enter well behind the apparent horizon. We have chosen $t_1=6$ and a) $\Delta t=0.4$ and b) $\Delta t=1.4$. In a dashed line we have plotted the apparent horizon}
\end{figure}

In the Schwinger-Keldysh or real time formalism, the thermal entropy can be understood as entanglement entropy in the pure state \eqref{skst} defined on two copies of the system, when tracing over one of them. In \cite{Maldacena:2001kr} the two copies of the system were put in correspondence with the two asymptotic regions of the extended Penrose diagram of the AdS eternal black hole. In this extended diagram there are geodesics that connect the two boundaries, reaching behind the horizon without falling into the singularity. Namely, those for which the constant C \eqref{CC} is positive and the signs \eqref{ee} satisfy $\epsilon\!=\!-1$ and ${\tilde \epsilon}\!=\!1$ (see Fig.\ref{fig:btz}). The function $\tilde s$ in the holographic model is derived from the length of geodesics whose endpoints have a large spatial separation, corresponding to small value of the integral of motion $r_\ast$. Relations \eqref{rl} and \eqref{longpar}, together with \eqref{trc2}, imply that from quite early on in the evolution the BTZ segments of these long geodesics are precisely of the mentioned type (see Fig.\ref{fig:behindhorizon} for examples). This offers a very consistent picture. The coarse grained entropy that describes the approach to thermalization requires information from behind the apparent horizon. Moreover  this information is extracted from geodesics that in the static black hole would join the two boundaries, strongly reminding to the real time formalism.

\begin{acknowledgments}
We thank J. Abajo-Arrastia, J. Barb\'on, P. Calabrese, J. Cardy, M. Garc{\'\i}a-Perez, K. Landsteiner, C. Pena, S. Ross, G. Sierra and S. Theisen for useful discussions. This work has been partially supported by grants FPA-2009-07908, HEPHACOS S2009/ESP-1473 and CPAN (CSD2007-00042). J. Apar{\'\i}cio is supported by the Portuguese Funda\c c\~ao para a Ci\^encia e Tecnologia, grant SFRH/BD/45988/2008.
\end{acknowledgments}

\begin{appendix}
\section{Geodesic trajectories}

Below can be found the explicit expressions for the different geodesic segments in the collapsing shell geometry. 
When $d_\text{in}\!=\!1$, the inner geodesic segment is given by
\be
\begin{array}{l}
r\in [ r_m, r_c ]  \;\;\; : \left\{ \begin{array}{l} {\hat v}(r)=t_0+t^{AdS}(r)-1/r \\ {\hat x}(r)=x^{AdS}(r) \end{array} \right. \\[5mm]
r\in [r_m, r_M]   \; : \left\{ \begin{array}{l} {\hat v}(r)=t_0-t^{AdS}(r)-1/r \\  {\hat x}(r)=-x^{AdS}(r)  \end{array} \right. 
\end{array} \;\;\; ,
\label{geoinp}
\ee
where $r_m$ given by \eqref{rmsp}, and $r_M\!=\!{\bar r}_c$ if there is a second crossing of the geodesic with the shell and infinity otherwise. 
The functions $t^{AdS}$ and $x^{AdS}$ are given in \eqref{functionsAdS}. When instead $d_\text{in}\!=\!-1$, we have
\be
r\in[r_c, r_M]  \; : \left\{ \begin{array}{l} {\hat v}(r)=t_0-t^{AdS}(r)-1/r \\  {\hat x}(r)=-x^{AdS}(r)  \end{array} \right. \, .
\label{geoinm}
\ee 
The integration constant $t_0$ has to be chosen such that $v_{d_{\rm in}}(r_c)\!=\!0$, and for simplicity we have set $x_0\!=\!0$ since it does not affect any relevant quantity.

The outer geodesic segment starting at the intersection point $r_c$ has the following expression. When $\eta\!=\!1$ we have
\be
r\in [r_c,\infty] \; : \left\{ \begin{array}{l}
v(r) = t_1-t_{d_{\rm out}}(\infty)+v_{d_{\rm out}}(r) \\
x(r) = x_+(r)-x_+(r_c)+d_{\rm in} x^{AdS}(r_c) \end{array} \right.\, ,
\label{eqin1}
\ee
with $t_d$ and $x_d$ given by \eqref{functions}. For $\eta\!=\!-1$ we must have $d_{\rm out}\!=\!-1$ and then
\be
\begin{array}{l}
r \in [r_c,r_m^{(1)}]  \;\, : \left\{ \begin{array}{l}
v(r) = t_1-t_+(\infty)+v_-(r) \\
x(r) = x_-(r)-x_-(r_c)+d_{\rm in} x^{AdS}(r_c) \end{array} \right. \\[5mm]
r \in [r_m^{(1)},\infty]  \; : \left\{ \begin{array}{l}
v(r) = t_1-t_+(\infty)+v_+(r) \\
x(r) = x_+(r)-x_-(r_c)+d_{\rm in} x^{AdS}(r_c) \end{array} \right. 
\end{array} \;\;\; ,
\label{eqin2}
\ee 
with the minimal value of the radial coordinate $r_m^{(1)}$ defined in \eqref{rmin}. 
The value of the time coordinate at the endpoint of this geodesic segment, $t_1$, is given by
\be
t_1=t_{\eta \, d_\text{out}}(\infty) - v_{d_\text{out}}(r_c) \, .
\label{t1}
\ee

The outer geodesic segment starting at the intersection point ${\bar r}_c$, when ${\bar \eta}\!=\!1$ is given by
\be
r\in [{\bar r}_c,\infty] \; : \left\{ \begin{array}{l}
{\bar v}(r) = t_2-{\bar t}_{{\bar d}_{\rm out}}(\infty)+{\bar v}_{{\bar d}_{\rm out}}(r) \\
{\bar x}(r) = {\bar x}_-(r)-{\bar x}_-({\bar r}_c)- x^{AdS}({\bar r}_c) \end{array} \right.\, ,
\ee
where the bar over the functions indicates that the energy parameter should be taken to be ${\bar A}_\text{out}$ instead of $A_\text{out}$.
For ${\bar \eta}\!=\!-1$ we have 
\be
\begin{array}{l}
r \in [{\bar r}_c,r_m^{(2)}]  \;\, : \left\{ \begin{array}{l}
{\bar v}(r) = t_2-{\bar t}_+(\infty)+{\bar v}_-(r) \\
{\bar x}(r) = {\bar x}_+(r)-{\bar x}_+({\bar r}_c)- x^{AdS}({\bar r}_c) \end{array} \right. \\[5mm]
r \in [r_m^{(2)},\infty]  \; : \left\{ \begin{array}{l}
{\bar v}(r) = t_2-{\bar t}_+(\infty)+{\bar v}_+(r) \\
{\bar x}(r) = {\bar x}_-(r)-{\bar x}_+({\bar r}_c)-x^{AdS}({\bar r}_c) \end{array} \right. 
\end{array} \;\;\; ,
\ee 
with 
\be
t_2={\bar t}_{{\bar \eta} \, {\bar d}_\text{out}}(\infty) - {\bar v}_{{\bar d}_\text{out}}({\bar r}_c) \, .
\label{t2}
\ee

From the previous expressions we can also read immediately $l$, the spatial distance between the geodesics endpoints. If there are two intersection points with the shell we have
\be
l=x_+(\infty)-x_\eta(r_c) +d_\text{in} x^{AdS}(r_c) +x^{AdS}({\bar r}_c) +{\bar x}_+(\infty)-{\bar x}_{{\bar \eta}}({\bar r}_c) \, ,
\label{ll}
\ee
where we have used that ${\bar x}_{-{\bar \eta}}({\bar r}_c) \!-\!{\bar x}_-(\infty)\!=\!{\bar x}_+(\infty)\!-\!{\bar x}_{{\bar \eta}}({\bar r}_c)$.
Otherwise
\be
l=x_+(\infty)-x_\eta(r_c) +d_\text{in} x^{AdS}(r_c) +x^{AdS}(\infty)  \, .
\ee

\section{Geodesics in the steady regime}

\begin{figure}[thbp]
\centering
\begin{tabular}{cc}
\includegraphics[width=7cm]{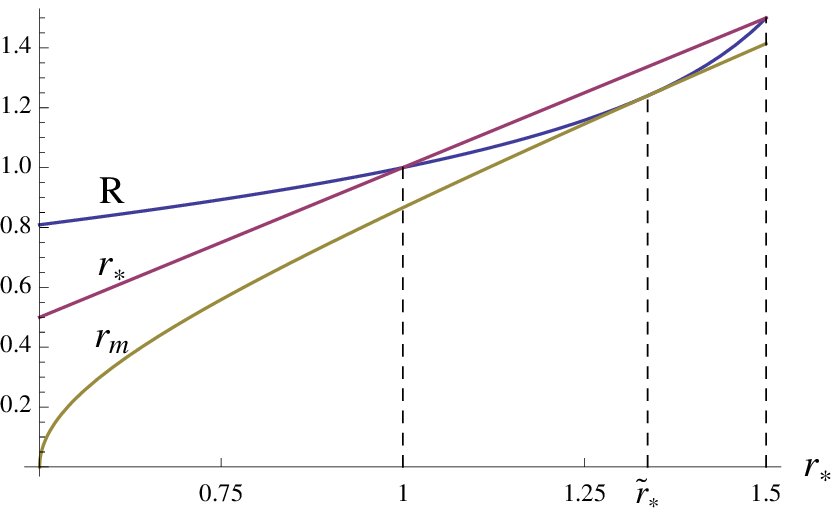} & \includegraphics[width=7cm]{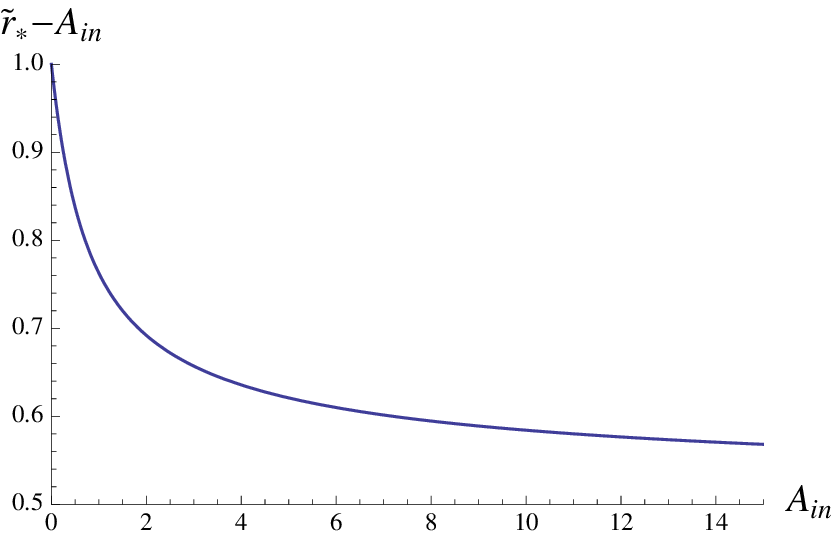}
\end{tabular}
\caption{\label{fig:R} For $r_h\!=\!1$, a) plot of $R$, $r_\ast$ and $r_m$ for $A_\text{in}\!=\!.5$; b) plot of ${\tilde r}_\ast (A_\text{in})\!-\!A_\text{in}$.}
\end{figure}

We will start deriving several properties of the function $R$ \eqref{r0ex}, which bounds the allowed values of the intersection point $r_c$ with the shell for geodesics that do not fall into the black hole singularity.  The function $R$ depends on $r_\ast$ and $A_\text{in}$. Its derivative with respect to $r_\ast$ is always positive, and thus it is an increasing function of this variable for fixed $A_\text{in}$. By substituting in \eqref{r0ex}, it is easy to verify that
\be
r_\ast=r_h,r_h+A_\text{in} \; \Rightarrow \; R=r_\ast \, .
\label{Rr}
\ee
These are the only points where this property is verified. Also easy to check is that
\be
{d R \over d r_\ast} \Big|_{r_h}={r_h \over r_h +A_\text{in}} <1 \, ,
\ee
which implies
\be
r_h<r_\ast < r_h+A_\text{in} \, \Rightarrow \; R<r_\ast  \, , \hspace{1cm} r_\ast<r_h \, \Rightarrow \; R>r_\ast \, .
\label{use2}
\ee
Between $r_h$ and $r_h\!+\!A_\text{in}$ there is a single ${\tilde r}_\ast$ at which $R\!=\!r_m$, with $r_m$ the minimal radial coordinate of spacelike AdS$_3$ geodesics.
The value of ${\tilde r}_c$ is given by
\begin{eqnarray} 
{\tilde r}_\ast&=&{1 \over 3} \left[ r_h \!+\! A_\text{in}\! +\! {2^{2/3}\big(r_h^2 \!+\! 2 A_\text{in} r_h \!+\! 4 A_\text{in}^2\big) \over  \Big(4r_h^3 \!-\! 15 A_\text{in}r_h^2\! -\! 24 A^2_\text{in}r_h\! -\! 32 A^3_\text{in} \!+\! 3 \sqrt{3} \sqrt{\!-8 A_\text{in}r_h^5 \!-\! 13 A^2_\text{in}r_h^4\! -\! 16 A^3_\text{in}r_h^3}\Big)^{1/3} } \right. \nonumber \\[5mm]
&& \;\;\;\; \left. + { \Big( 4r_h^3\! -\! 15 A_\text{in}r_h^2 \!-\! 24 A^2_\text{in}r_h \!-\! 32 A^3_\text{in}\! +\! 
   3 \sqrt{3} \sqrt{\!-8 A_\text{in}r_h^5\! -\! 13 A^2_\text{in}r_h^4 \!-\! 16 A^3_\text{in}r_h^3} \Big)^{1/3}\over 2^{2/3} } \,  \right]   ,
\label{rslimit}
\end{eqnarray}
The fact that $r_\ast\!>\!r_m$ together with \eqref{Rr}, imply that $R$ is bigger than $r_m$ for any other $r_\ast$. All these properties are resumed in Fig.\ref{fig:R}a.
In Fig.\ref{fig:R}b we have plotted ${\tilde r}_\ast\!-\!A_\text{in}$. This is a decreasing function of the energy parameter that coincides with the horizon radius at vanishing $A_\text{in}$ and tends to $r_h/2$ for large values of $A_\text{in}$.

We will study now the properties of geodesics whose intersection points with the infalling shell are close to $R$. By direct inspection, one can check that the parameter $C$ \eqref{CC} vanishes at $r_c\!=\!R$, implying that   
\be
r_\ast^2-A_\text{out}^2+r_h^2\big|_R=2{\tilde \epsilon}(R) r_\ast r_h\, .
\label{RC0}
\ee
Since $A_\text{out}^2$ \eqref{aout} is a continuous function of $r_\ast$ over the whole interval $[A_\text{in}, r_h\!+\!A_\text{in}]$, ${\tilde \epsilon}(R)$ must take a unique value. We can therefore check it at a particular $r_\ast$, which we conveniently choose to be the upper limit of the interval. At $r_\ast\!=\!r_h\!+\!A_\text{in}$ the inner and outer energy parameters coincide, $A_\text{out}\!=\!A_\text{in}$, and hence ${\tilde \epsilon}(R)\!=\!1$. Substituting back in \eqref{RC0}, when $r_c\!\rightarrow\!R$ we then have
\be
A_\text{out}=|r_h-r_\ast| +\sign(r_h-r_\ast) f \delta r_{c} \, ,
\label{Alimit2}
\ee
where $\delta r_{c}\!=\!|r_{c}\!-\!R|/R$ and $f$ is a function 
whose explicit expression we will not need.
In the same limit we can check that
\be
\epsilon=-\sign(r_h-r_\ast) \, .
\ee

Repeating the continuity argument above shows that the function whose modulus defines $A_\text{out}$ verifies at $r_c\!=\!R$ 
\be
{A_\text{in}(2 R^2-m)- d_\text{in} m \sqrt{A_\text{in}^2+R^2-r_\ast^2}  \over 2 R^2} =\pm(r_\ast-r_h) \, ,
\ee
where one sign on the rhs must apply to the complete interval where $R$ is defined. Selecting again $r_\ast\!=\!r_h\!+\!A_\text{in}$, we easily obtain that the correct sign is plus.
Substituting this relation in \eqref{dout} we derive after some algebra
\be
d_\text{out}=-\sign\big( (m-2R^2)(r_h+A_\text{in}-r_\ast)+m A_\text{in} \big) \, .
\ee
It can be checked that the zeros as well as the sign of the previous function coincide with those of the simpler function $R^2\!-\!r_h r_\ast$, and hence
 \be
d_\text{out}=-\sign( R^2-r_h r_\ast) \, .
\ee
Finally, using \eqref{etaa} and the properties of $R$ summarized in Fig.\ref{fig:R}a we deduce 
\be
\eta \, d_\text{out}=1 \, .
\ee
Analogous expressions hold for the second crossing whenever ${\bar r}_c\!\simeq\!R$. 

The previous results provide all the necessary ingredients to derive the values of $t_i$, $l$ and $L_{reg}$ in the limit that the intersection points with the shell are close to $R$. We start by analyzing the value of the time coordinate at the geodesic endpoint $t_1$, given by \eqref{t1}. Using several of the results above, we get 
\be
t_{\eta d_\text{out}} (\infty)= -{1 \over 2 r_h} \log \big( |r_\ast^2-(r_h \sign(r_h\!-\!r_\ast)\!-\! A_{\text{out}})^2 | m^{-1} \big) \, .
\label{tinf2}
\ee
From \eqref{Alimit2} we see that the argument of the logarithm vanishes as $r_c$ tends to $R$. However $v_{d_\text{out}}(r_c)$, also contributing to $t_1$, remains finite in that limit
\be
v_{d_\text{out}}(R)=-{1 \over 2 r_h} \log {4 \big| (r_h-r_\ast)(R^2-r_h r_\ast) \big| \over r_h (r_h-R)^2} \, .
\ee
As explained in Section \ref{sec:latet}, in the limit that both intersections points are close to $R$, the energy parameter of the inner geodesic segment $A_\text{in}$ must be small and of the order of $\delta r_c$. We are interested in obtaining the value of $t_1$ up to terms that vanish in the limit $r_c,{\bar r_c}\!\rightarrow\!R$. Therefore it makes sense to set $A_\text{in}\!=\!0$ in the finite contributions to $t_1$. Combining the previous two pieces with this further approximation, we obtain
\be
t_1= -{1 \over 2 r_h} \log \delta r_c +{1 \over 2 r_h} \log \left| {2 r_h (r_h-r_\ast)(R_0^2-r_h r_\ast)  \over r_\ast f_0 (r_h-R_0)^2} \right| \, ,
\label{ft}
\ee
where $R_0$ and $f_0$ are the values of $R$ and $f$ for $A_\text{in}\!=\!0$. The same expression applies for $t_2$ replacing $\delta r_c$ by  $\delta {\bar r}_c$. 

Several pieces contribute to the spatial separation $l$ of the geodesic endpoints, given by \eqref{ll}. From the outer BTZ geodesic segment starting at $r_c$ we have:
\be
x_+(\infty)=-{1 \over 2 r_h} \log \big( | A_{\text{out}}^2-(r_h\!-\!r_\ast)^2 | m^{-1} \big) \, ,
\label{xinf2}
\ee
which is again divergent as $r_c$ tends to $R$, and $x_\eta(r_c)$, which remains finite
\be
x_\eta(R)=-{1 \over 2 r_h} \log {4|R^2-r_h r_\ast| r_\ast \over r_h R^2} \, .
\ee
There are analogous pieces from the crossing at ${\bar r}_c$. The geodesic segment inside the shell contributes to $l$ as $d_\text{in} x^{AdS}(r_c) +x^{AdS}({\bar r}_c)$, with $x^{AdS}$ given in \eqref{functionsAdS2}. This quantity is finite when the intersection points tend to $R$ and, as explained above, we can set $A_\text{in}\!=\!0$ to the order we are working. When $A_\text{in}\!=\!0$ the minimal value of the radial coordinate attained by the inner geodesic equals $r_\ast$ and, as seen in Fig.\ref{fig:R}b, ${\tilde r}_\ast\!=\!r_h$ in that case. The range of allowed data for the geodesics \eqref{rangeint} implies then that $r_c$ can only happen in the branch $d_\text{in}\!=\!1$. Therefore the contribution of the inner geodesic to $l$ is $2 x^{AdS}(R_0)$.
Collecting all these pieces we obtain
\be
l=-{1 \over 2 r_h} \log (\delta r_{c} \delta {\bar r}_{c} )  + {1 \over 2 r_h} \log \left| {2 r_h r_\ast (R_0^2-r_h r_\ast)  \over (r_h-r_\ast) f_0 R_0^2} \right|+ {4 (r_h-r_\ast) R_0 \over m r_\ast}  \, .
\label{fl}
\ee

The geodesic length gets a divergent contribution from the piece $\log C$ in \eqref{geolenbtz} in the limit in which the intersection points tend to $R$, given that in this limit $C$ vanishes. Setting further $A_\text{in}\!=\!0$ and substituting in \eqref{geolenads} and \eqref{geolenbtz}, we have
\be
L_{reg}= -{1 \over 2} \log (\delta r_{c} \delta {\bar r}_{c} )  + \log \left| {2 (R_0^2-r_h r_\ast) \over f_0 r_h r_\ast (r_h-r_\ast)} \right| + 2\log {R_0 + \sqrt{R_0^2-r_\ast^2} \over r_\ast} \, .
\label{fL}
\ee  

\end{appendix}

\bibliographystyle{JHEP}
\bibliography{therm}
\end{document}